\documentclass[aps,prl,twocolumn,groupedaddress]{revtex4-1}

\usepackage{graphicx}
\usepackage{enumitem} 
\usepackage{multirow} 
\usepackage{array} 

\newcolumntype{L}[1]{>{\raggedright\let\newline\\\arraybackslash\hspace{0pt}}m{#1}} 
\newcolumntype{C}[1]{>{\centering\let\newline\\\arraybackslash\hspace{0pt}}m{#1}} 
\newcolumntype{R}[1]{>{\raggedleft\let\newline\\\arraybackslash\hspace{0pt}}m{#1}} 

\begin{document}

\title{Inequity exemplified by underrepresented minority students receiving lower grades and having higher rates of attrition across STEM disciplines
}

\author{Kyle~M.~Whitcomb}
\affiliation{Department of Physics and Astronomy, University of Pittsburgh, Pittsburgh, PA, 15260}
\author{Chandralekha~Singh}
\affiliation{Department of Physics and Astronomy, University of Pittsburgh, Pittsburgh, PA, 15260}

\date{\today}

\begin{abstract}
	Underrepresented minority (URM) students are subjected to historically rooted inequities when pursuing an education, especially in STEM disciplines with little diversity.
	In order to make STEM education equitable and inclusive, evidence for how students from different racial/ethnic demographics are faring is necessary.
	We use 10 years of institutional data at a large public university to investigate trends in the majors that Asian, URM, and White students declare, drop after declaring, and earn degrees in as well as the GPA of the students who drop or earn a degree.
	We find that higher percentages of the URM students drop most majors compared to other students and these trends are particularly pronounced in physics and economics.
	Moreover, we find alarming GPA trends in that the URM students consistently earn lower grades than their Asian and White peers.
	Furthermore, in some STEM disciplines, the URM students who earn a degree are earning the same grades as the Asian and White students who dropped the major. This troubling trend may signify lack of sufficient support, mentoring, and guidance to ensure excellence of the URM students who are already severely disadvantaged.
	These quantitative findings call for 
	making learning environments equitable and inclusive so that many URM students who come to college with severe disadvantages are appropriately supported and can excel similar to other students.
\end{abstract}

\maketitle

\section{Introduction and Theoretical Framework}
\label{intro}

Increasingly, Science, Technology, Engineering, and Mathematics (STEM) departments across the US are focusing on using evidence to improve the learning of all students, regardless of their background and making learning environments equitable and inclusive~\cite{parappilly2019, johnson2012, johnson2017, mayo2009, metcalf2018, molin2019, king2016, maltese2011, maltese2017, means2018, gregorcic2018, borrego2008, borrego2011, borrego2014, johansson2018, henderson2008, henderson2012, dancy2010, bosser2019}.
However, racial/ethnic minority students are still severely underrepresented in many STEM disciplines~\cite{nsf2018}`
In order to understand the successes and shortcomings of the current state of education, the use of institutional data to investigate past and current trends is crucial.
In the past few decades, institutions have been keeping increasingly large digital databases of student records.
We have now reached the point where there are sufficient data available at many institutions for robust statistical analyses using data analytics that can provide invaluable information for transforming learning environments and making them more equitable and inclusive for all students~\cite{baker2014, papamitsiou2014}.
Studies utilizing many years of institutional data can lead to analyses that were previously limited by statistical power.
This is particularly true for studies of performance and persistence in STEM programs that rely on large sample sizes~\cite{ohland2008, lord2009, eris2010, maltese2011, min2011, lord2015, ohland2016, matz2017, witherspoon2019, king2016, safavian2019, maltese2017, means2018}.

In this study, we use 10 years of institutional data from a large state-related research university to investigate how patterns of student major-declaration and subsequent degree-earning may differ by self-reported race/ethnicity.
We will consider the two racial/ethnic groups that are over-represented in US Bachelor's degree attainment~\cite{nsf2018}, namely Asian and White students, separately from other students whom we call underrepresented minority (URM) students.
The theoretical framework for this study has two main foundations: critical theory and expectancy value theory.

Critical theories, e.g., of race and gender, focus on historical sources of inequities within society, i.e., societal norms that perpetuate obstacles to the success of certain groups of disadvantaged people~\cite{gopalan2019, crenshaw1995, kellner2003, yosso2005, gutierrez2009, taylor2009, tolbert2018, schenkel2020, metcalf2018}.
Critical theory tells us that the dominant group in a society perpetuates these norms, which are born out of their interests, and pushes back against support systems that seek to subvert these norms~\cite{crenshaw1995, kellner2003, yosso2005}.
In our case, critical race theory provides a historical perspective on the much-studied racial/ethnic inequities in STEM.
Much important work has been done that relates to critical theory of race and ethnicity in STEM education~\cite{ashby2018, wong2016, bancroft2018, johnson2017, solorzano2000, lewis2009, bang2010, ong2018, tolbert2018, green2019, mutegi2019, sheth2019, schenkel2020}, e.g., pervasive stereotypes about who can and cannot succeed in STEM puts additional burden on URM students in STEM courses.

Expectancy value theory (EVT) is another framework that is central to our investigation and states that a student's persistence and engagement in a discipline are related to their expectancy about their success as well as how the student values the task~\cite{eccles1984, eccles1990, eccles1994}.
In an academic context, ``expectancy,'' which refers to the individual's beliefs about their success in the discipline, is closely related to Bandura's construct of self-efficacy, defined as one's belief in one's capability to succeed at a particular task or subject~\cite{bandura1991, bandura1994, bandura1997, bandura1999, bandura2001, bandura2005, eccles1984, eccles1990, eccles1994}.

There are four main factors that influence students' expectancy or self-efficacy, namely vicarious experiences (e.g., instructors or peers as role models), social persuasion (e.g., explicit mentoring, guidance, and support), level of anxiety, and performance feedback (e.g., via grades on assessment tasks)~\cite{bandura1991, bandura1994, bandura1997, bandura1999, bandura2001, bandura2005}.
URM students generally have lower self-efficacy than non-URM students in many STEM disciplines because these four factors negatively influence them~\cite{astin1993, cross1993, bianchini2002, britner2006, bianchini2013, basile2015, hilts2018}.
For example, URM students are underrepresented in their classrooms across STEM, and less likely to have a role model they can relate to among the faculty~\cite{nsf2018}.
Further, the stereotypes surrounding URM students in many STEM disciplines can affect how they are treated by mentors, even if such an effect is subconscious~\cite{astin1993, cross1993, bianchini2002, britner2006, bianchini2013, basile2015, hilts2018}.
Moreover, URM students are susceptible to stress and anxiety from stereotype threat (i.e., the fear of confirming stereotypes about URM students in many STEM disciplines) which is not experienced by their non-URM peers~\cite{astin1993, cross1993, bianchini2002, britner2006, bianchini2013, basile2015, hilts2018}.
This stress and anxiety can rob them of their cognitive resources, especially during high-stakes assessments such as exams.

Expectancy can influence grades earned as well as the interest and likelihood to persist in a program~\cite{bandura1991, bandura1994, bandura1997, bandura1999, bandura2001, bandura2005, wan2019}.
Stereotype threats that URM students in many STEM disciplines experience can increase anxiety in learning and test-taking situations and lead to deteriorated performance.
Since anxiety can increase when performance deteriorates, these factors working against URM students in STEM can force them into a feedback loop and hinder their performance further, which can further lower their self-efficacy and can continue to affect future performance~\cite{bandura1991, bandura1994, bandura1997, bandura1999, bandura2001, bandura2005, wan2019}.

In EVT, value is typically defined as having four facets: intrinsic value (i.e., interest in the task), attainment value (i.e., the importance of the task for the student's identity), utility value (i.e., the value of the task for future goals such as career), and cost (i.e., opportunity cost or psychological effects such as stress and anxiety)~\cite{eccles1984, eccles1990, eccles1994}.
In the context of URM students' enrollment and persistence in many STEM disciplines, the societal stereotypes can influence all facets of the students' value of these STEM disciplines.
Intrinsic value can be informed by societal stereotypes surrounding the STEM disciplines, and attainment and utility values can be further tempered by these stereotypes.
Utility value is an important facet of student education in STEM, since a degree in a STEM field provides many job opportunities for graduating students.
In addition, the psychological cost of majoring in these disciplines can be inflated by the stereotype threat.
All of these effects can conspire to suppress the likelihood of students choosing and/or persisting in various STEM disciplines~\cite{wan2019}.

In order to measure the long-term effects of these systemic disadvantages, we investigate the differences in attrition rates and choices of major of Asian, URM, and White students over the course of their studies at one large public research university using 10 years of institutional data.
Since these disadvantages to students can be context-dependent, we will consider the attrition rates in many different STEM majors and non-STEM majors in order to understand the trends in each discipline.

\subsection{Research Questions}
\label{sec:rq}

Our research questions regarding the relationships between race/ethnicity and persistence in a degree are as follows.

\begin{enumerate}[label={\bfseries RQ\arabic*.}, ref={\bfseries RQ\arabic*}, itemsep=1pt]
	\item \label{rq_declare} How many students major in each discipline? How many Asian, URM, and White students major in each discipline?
	\item \label{rq_drop} Do rates of attrition from the various majors differ? Do rates of attrition from the various majors differ for Asian, URM, and White students?
	\item \label{rq_droppers} Among those students who drop a given major, what degree, if any, do those students earn? How do these trends differ for Asian, URM, and White students?
	\item \label{rq_degree} What fraction of declared majors ultimately earn a degree in that major in each STEM subject area? How do these trends differ for Asian, URM, and White students?
	\item \label{rq_gpa} What are the GPA trends over time among students who earn a degree in a given major and those who drop that major? How do these trends differ for Asian, URM, and White students?
\end{enumerate}

\section{Methodology} 
\label{methodology}

\subsection{Sample}
\label{sec:sample}

Using the Carnegie classification system, the university at which this study was conducted is a public, high-research doctoral university, with balanced arts and sciences and professional schools, and a large, primarily residential undergraduate population that is full-time and reasonably selective with low transfer-in from other institutions~\cite{carnegie}.

The university provided for analysis the de-identified institutional data records of students with Institutional Review Board approval.
In this study, we examined these records for $N = 18,009$ undergraduate students enrolled in two schools within the university: the School of Engineering and the School of Arts and Sciences.
This sample of students includes all of those from ten cohorts who met several selection criteria, namely that the student had first enrolled at the university in a Fall semester from Fall 2005 to Fall 2014, inclusive, and the student had either graduated and earned a degree, or had not attended the university for at least a year as of Spring 2019.
This sample of students is 49.8\% female and had the following race/ethnicities: 77.8\% White, 11.1\% Asian, 6.9\% Black, 2.5\% Hispanic, and 1.8\% other or multiracial.

\subsection{Measures}
\label{sec:measures}

\subsubsection{Race/Ethnicity}

The institutional data provided by the university included self-reported race/ethnicity of the students.
Students were asked to indicate all of the following racial/ethnic groups with which they identified: American Indian/Alaskan, Asian, Black, Hawaiian/Pacific Islander, Hispanic, White, or Other.
The researchers converted the student answers to this question into three categories for analysis, defined as follows.
\begin{itemize}
	\item ``Asian'': Students who selected only Asian or only Asian and White.
	\item ``White'': Students who selected only White.
	\item ``URM'': Underrepresented minority students who selected any option(s) other than Asian or White. 
\end{itemize}
A very small number of students chose not to answer this question, and were removed from analysis. 

\subsubsection{Academic Performance}

Measures of student academic performance were also included in the provided data.
High school GPA was provided by the university on a weighted scale from 0-5 that includes adjustments to the standard 0-4 scale for Advanced Placement and International Baccalaureate courses.
The data also include the grade points earned by students in each course taken at the university.
Grade points are on a 0-4 scale with $\textrm{A}=4$, $\textrm{B}=3$, $\textrm{C}=2$, $\textrm{D}=1$, $\textrm{F}=0$, where the suffixes ``$+$'' and ``$-$'' add or subtract, respectively, $0.25$ grade points (e.g. $\textrm{B}-=2.75$), with the exception of $\textrm{A}+$ which is reported as the maximum 4 grade points.
The courses were categorized as either STEM or non-STEM courses, with STEM courses being those courses taken from any of the following departments: biological sciences, chemistry, computer science, economics, any engineering department, geology and environmental science, mathematics, neuroscience, physics and astronomy, and statistics.
We note that for the purposes of this paper, ``STEM'' does not include the social sciences other than economics, which has been included due to its mathematics-intensive content.

\subsubsection{Declared Major and Degree Earned}

For each student, the data include their declared major(s) in each semester as well as the major(s) in which they earned a degree, if any.
The data were transformed into a set of binary flags for each semester, one flag for each possible STEM major as well as psychology and a general non-STEM category for all other majors.
A similar set of flags was created for the degrees earned by students.
From these flags, we tabulated a number of major-specific measures in each semester, including
\begin{itemize}
	\item current number of declared majors,
	\item number of newly declared majors from the previous semester,
	\item number of dropped majors from the previous semester,
	\item number of retained majors from the previous semester.
\end{itemize}
The total number of unique students that ever declared or dropped a major were also computed.
The subset of students that dropped each major were further investigated and the major in which they ultimately earned a degree, if any, was determined.

Throughout this paper we group the STEM majors into two clusters: chemistry, computer science, engineering, mathematics, and physics and astronomy; and biological sciences, economics, geology and environmental science, neuroscience, and statistics.
When ordering majors (i.e., in figures and tables), the majors will be presented in the order they are listed in the previous sentence (first by group, then alphabetically within each group), followed by non-STEM and psychology.
Further, we group the final three STEM majors (geology and environmental science, neuroscience, and statistics) into a category labeled ``Other STEM'' for figures and tables.
Similarly, ``engineering'' groups together all engineering majors for departments in the School of Engineering at the studied university.
These majors include chemical, computer, civil, electrical, environmental, industrial, and mechanical engineering as well as bioengineering and materials science.

Finally, we will make use of shortened labels for the majors in figures and tables.
These shortened labels are defined in Table~\ref{table_major_labels}.

\begin{table}
	\centering
	
	\begin{tabular}{l l}
		Major & Short Label \\
		\hline
		Chemistry				& Chem \\
		Computer Science		& CS \\
		Engineering				& Engr \\
		Mathematics				& Math \\
		Physics and Astronomy	& Phys \\
		Biological Sciences		& Bio \\
		Economics				& Econ \\
		Geology and				& \multirow{2}{*}{Other STEM} \\
		Environmental Science	& \\
		Neuroscience			& Other STEM \\
		Statistics				& Other STEM \\
		Non-STEM				& Non-STEM \\
		Psychology				& Psych
	\end{tabular}
	
	\caption{\label{table_major_labels}
		A list of the majors considered in this study and the shortened labels used to refer to those majors in tables and figures.
	}
\end{table}

\subsubsection{Year of Study}

Finally, the year in which the students took each course was calculated from the students' starting term and the term in which the course was taken.
Since the sample only includes students who started in fall semesters, each ``year'' contains courses taken in the fall and subsequent spring semesters, with courses taken over the summer omitted from this analysis.
For example, if a student first enrolled in Fall 2007, then their ``first year'' occurred during Fall 2007 and Spring 2008, their ``second year'' during Fall 2008 and Spring 2009, and so on in that fashion.
If a student is missing both a fall and spring semester during a given year but subsequently returns to the university, the numbering of those post-hiatus years is reduced accordingly.
If instead a student is only missing one semester during a given year, no corrections are made to the year numbering.

\subsection{Analysis}
\label{sec:analysis}

For each student, we calculate their grade point average (GPA) across courses taken in each year of study from their first to sixth years.
In addition, we calculate the student's STEM GPA in each year, that is, the GPA in STEM courses alone.
The mean GPA as well as the standard error of the mean is computed for various groupings of students~\cite{freedman2007}.

Further, proportions of students in various groups (i.e., grouped by major and/or demographic group) are calculated along with the standard error of a proportion~\cite{freedman2007}.
In particular, the proportions we report are
\begin{itemize}
	\item the proportion of students in each major that are Asian, URM, or White students,
	\item the proportion of Asian, URM, and White students, respectively, that declare each subject as a major,
	\item the proportion of declared majors that drop the major,
	\item the proportion of those who drop each major that earn a degree in each other major, and
	\item the proportion of all declared majors that ultimately earn a degree in that major.
\end{itemize}

All analyses were conducted using R~\cite{rcran}, making use of the package \texttt{tidyverse}~\cite{tidyverse} for data manipulation and plotting.

\section{Results}
\label{results}

\subsection{Major Declaration Patterns}
\label{sec:major_patterns}

\begin{figure*}
    \centering
    
	\includegraphics[width=0.95\textwidth]{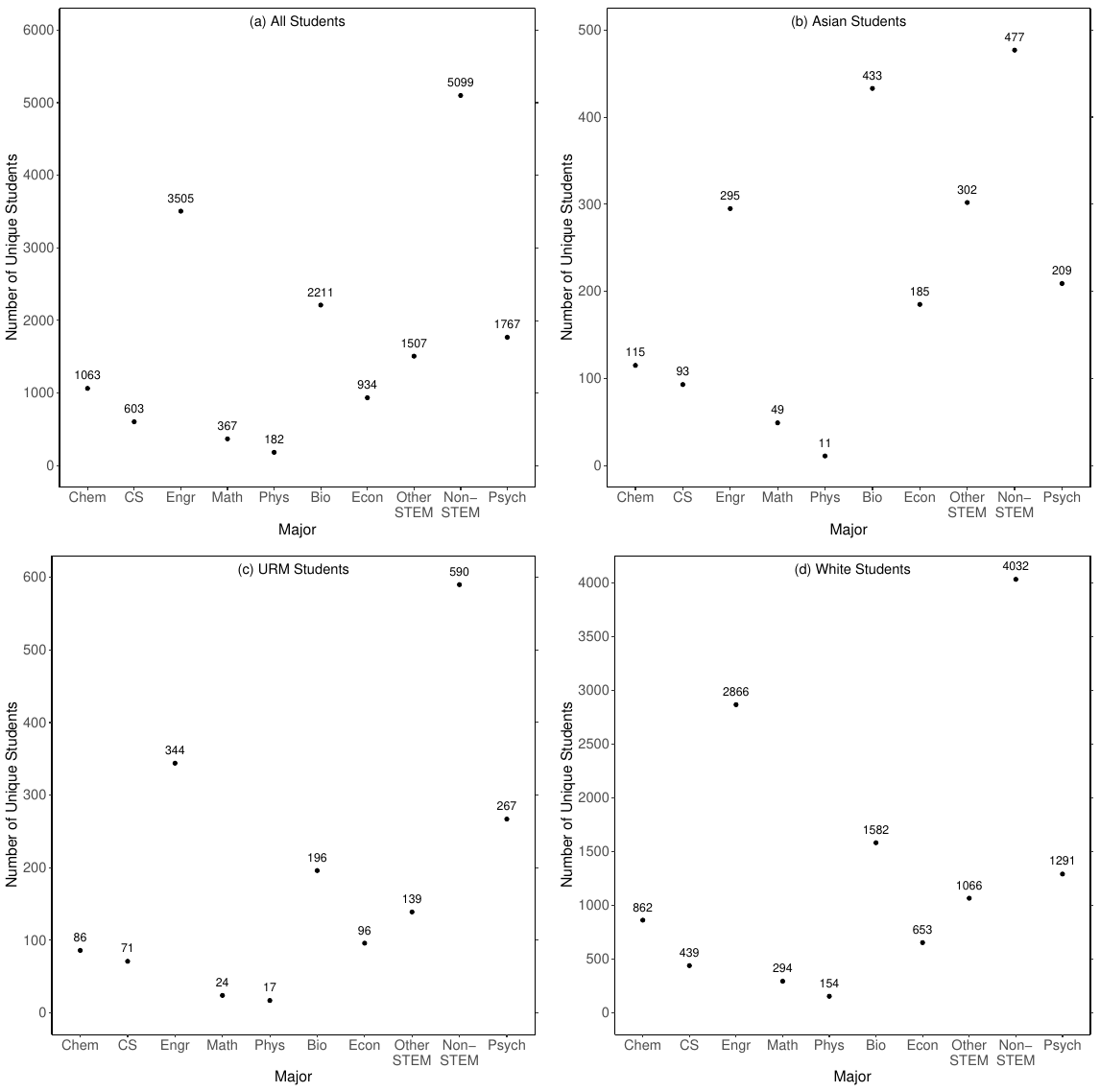}
	
	\caption{\label{figure_n_unique}
		For each major on the horizontal axis, the number of unique students in the sample that ever declared that major is listed.
		Since students may change majors or declare multiple majors, some students may contribute to the counts of more than one major.
		These counts are calculated separately for (a) all students, (b) Asian students, (c) URM students, and (d) White students.
		Note that the scale of the vertical axes differs for each plot.
		}
\end{figure*}

There are many angles with which we can approach \ref{rq_declare} and investigate patterns of student major declaration.
First, Fig.~\ref{figure_n_unique} shows the number of students that ever declared each major.
This is done both overall (Fig.~\ref{figure_n_unique}a) and for Asian students (\ref{figure_n_unique}b), URM students (\ref{figure_n_unique}c), and White students (\ref{figure_n_unique}d) separately.
These results provide an important context for the upcoming analyses that may be partially explained by the number of students in each major.

\begin{figure*}
    \centering
    
	\includegraphics[width=0.95\textwidth]{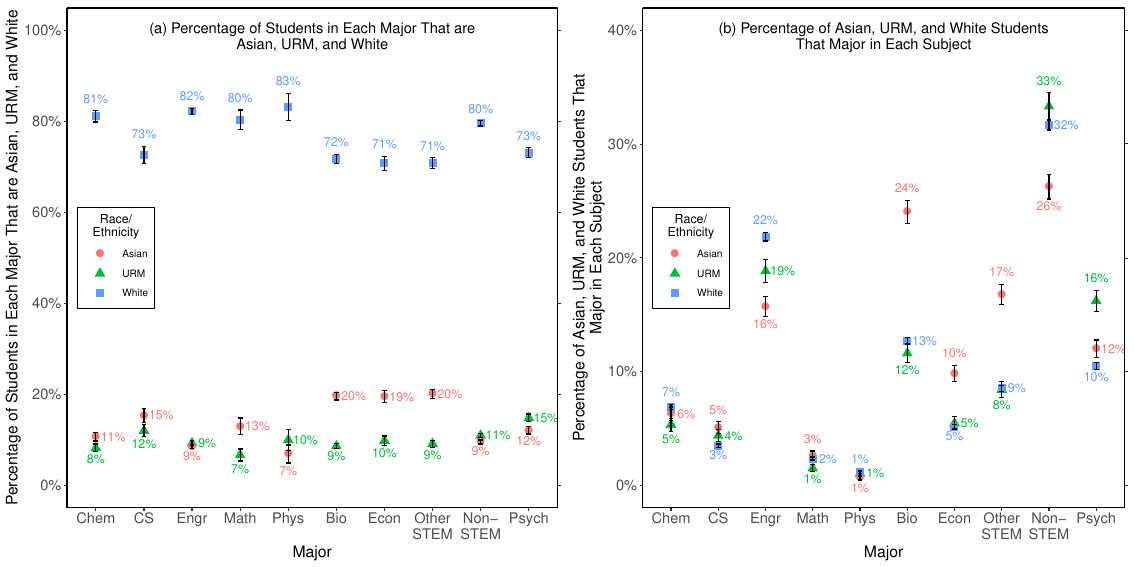}
	
	\caption{\label{figure_percentages}
		In (a), the percentages of students in each major that are Asian, URM, or White students are calculated (i.e., the percentages in each column will sum to roughly 100\%).
		In (b), the percentages of students in each racial/ethnic group that major in each subject are calculated (i.e., the percentages in each group will sum to roughly 100\%).
		Discrepancies in the sum of percentages may occur due to rounding the listed percentages to the nearest integer as well as, in (b), students declaring multiple majors.
		}
\end{figure*}

Figures~\ref{figure_n_unique}b and \ref{figure_n_unique}c begin to hint at differences in enrollment patterns for the different racial/ethnic groups.
URM and White students (Figs.~\ref{figure_n_unique}c and \ref{figure_n_unique}d) have relatively similar patterns, with some small differences such as a slightly higher fraction of White students declaring engineering majors and a slightly higher fraction of URM students declaring psychology majors.
Asian students show some trends that differ from both URM and White students, in particular a higher fraction of Asian students major in biological sciences, other STEM (i.e., geology and environmental science, neuroscience, and statistics), and economics.
These patterns are explored further in Fig.~\ref{figure_percentages} by standardizing the scales in two ways.
In Fig.~\ref{figure_percentages}a, we consider the populations of each major separately and calculate the percentages of that population that consist of Asian, URM, or White students.
This provides insight into what these students might be seeing in the classes for their major.
The trends seen in Fig.~\ref{figure_percentages}a are largely dominated by the relative representation of Asian, URM, and White students in the university as a whole.

Another way to represent the population of these majors is to consider what percentage of all Asian, URM, or White students choose each major, as seen in Fig.~\ref{figure_percentages}b.
While the main signal of this plot mimics those of Figs.~\ref{figure_n_unique}b,  \ref{figure_n_unique}c, and \ref{figure_n_unique}d, we can now see the differences noted earlier more clearly.
For example, in Fig.~\ref{figure_percentages}b we can again see that the majors chosen by URM and White students are similar, with the exceptions of engineering (chosen by slightly more White students) and psychology (chosen by slightly more URM students).
We can also see in Fig.~\ref{figure_percentages}b the differing trends for Asian students noted earlier, with Asian students more often choosing majors in biological sciences, economics, and the other STEM category, and now we can see that conversely Asian students are less likely than URM and White students to major in engineering and non-STEM.

\begin{figure*}
    \centering
    
	\includegraphics[width=0.95\textwidth]{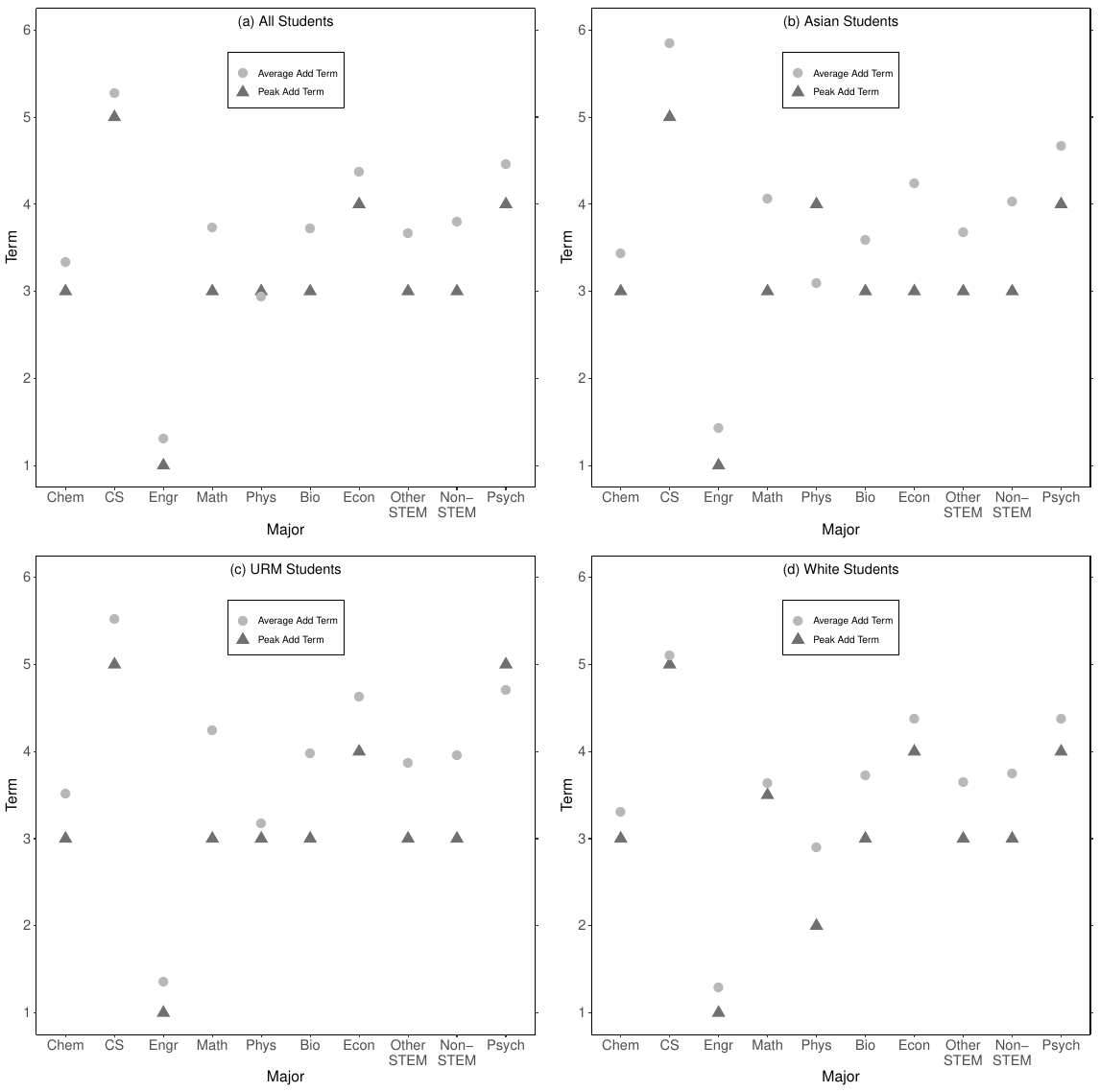}
	
	\caption{\label{figure_peak_avg_term}
	For each major, the term with the peak number of students adding the major in that term is plotted (triangles) as well as the average term in which students add that major (circles).
	This is done separately for (a) all students, (b) Asian students, (c) URM students, and (d) White students.
	Note that the value of 3.5 for Math in (d) indicates that an equal number of White students added a math major in terms 3 and 4.}
\end{figure*}

Finally, another piece of information about enrollment patterns that is missing from Figs.~\ref{figure_n_unique} and \ref{figure_percentages} is when these students declare each major.
Figure~\ref{figure_peak_avg_term} shows, for each major, the average term in which students added the major as well as the peak term (that is, the term with the highest number of new students adding the major).
As with Fig.~\ref{figure_n_unique}, this is done separately for all students (Fig.~\ref{figure_peak_avg_term}a), Asian students (\ref{figure_peak_avg_term}b), URM students (\ref{figure_peak_avg_term}c), and White students (\ref{figure_peak_avg_term}d).

For the majority of majors in Fig.~\ref{figure_peak_avg_term}a, the peak of students adding the major is in the third term (that is, the start of their second year), with an average of three to four terms.
Two majors, economics and psychology, depart slightly from this general trend, each with a peak in the fourth term and an average between fourth and fifth terms.
Two other majors, computer science and engineering, depart more significantly from the general trend in ways that can be explained by their particular implementation at the studied university.

Engineering has a peak in the first term in Fig.~\ref{figure_peak_avg_term}a, with an average only slightly later.
Engineering is the earliest major in our data, since all admitted students who enroll in the School of Engineering are considered ``undeclared engineering'' majors (they have not declared subdisciplines within engineering) and so the majority of engineering students are in the engineering program in their first term.
Computer science instead has the latest peak term in Fig.~\ref{figure_peak_avg_term}a, namely in the fifth term with a slightly higher average.
This is due to the structure of the computer science program at the studied university, which does not allow students to declare the major until they have completed five of the required courses for the major.
These trends in engineering and computer science are important to keep in mind while considering the results presented elsewhere in this paper, since in computer science we are not able to capture attrition that occurs during the terms before a student officially declares a major.
Conversely in engineering, we are able to capture almost all attrition in the first year due to the unique enrollment conditions of engineering students, which is also not possible for majors within the School of Arts and Sciences who choose majors at their convenience.

Turning then to Figs.~\ref{figure_peak_avg_term}b, \ref{figure_peak_avg_term}c and \ref{figure_peak_avg_term}d, we see almost identical trends as in Fig.~\ref{figure_peak_avg_term}.
Broadly, the differences occur for majors in which two terms have similar numbers of students declaring that major.
When considering a subset of all students, which of these two terms is the peak will sometimes shift.

A more detailed accounting of the number of students that enroll in each term for each major are reported in Tables~\ref{table_adddrop_1} and \ref{table_adddrop_2} in Appendix A.
Further, summaries of the total number of unique students as well as the peak term and number of concurrent students in each major, students adding each major, and students dropping each major are available in Tables~\ref{table_appendix_all_main}, \ref{table_appendix_asian_main}, \ref{table_appendix_urm_main}, and \ref{table_appendix_white_main} in Appendix B.

\subsection{Attrition Rates}
\label{sec:attrition}

\begin{figure*}
    \centering
    
	\includegraphics[width=0.95\textwidth]{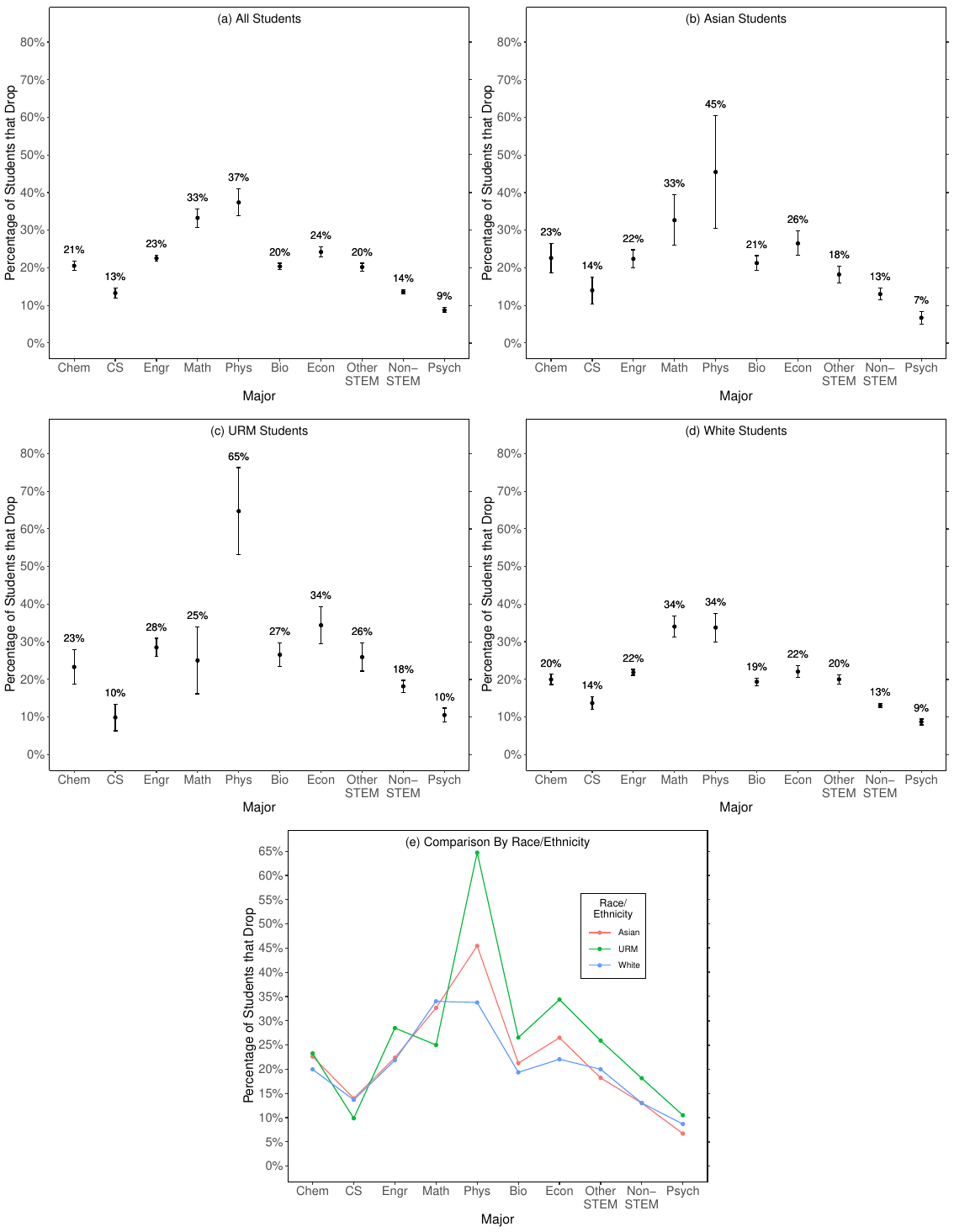}
	
	\caption{\label{figure_drop_plots}
		For each major, the percentage of students who declared the major but subsequently dropped the major is plotted along with its standard error.
		This is done separately for (a) all students, (b) Asian students, (c) URM students, and (d) White students.
		The plots for the different racial/ethnic groups are combined into a single plot (e) with the error bars omitted for visibility and ease of comparison, along with lines connecting different points as guides to the eye.
		}
\end{figure*}

In order to answer \ref{rq_drop}, we further considered patterns of attrition rates by race/ethnicity.
In Fig.~\ref{figure_drop_plots}, we consider the drop rates of students in each major or group of majors for all students (Fig.~\ref{figure_drop_plots}a), Asian students (Fig.~\ref{figure_drop_plots}b), URM students (Fig.~\ref{figure_drop_plots}c), and White students (Fig.~\ref{figure_drop_plots}d).
Further, Figs.~\ref{figure_drop_plots}b, \ref{figure_drop_plots}c, and \ref{figure_drop_plots}d are combined into Fig.~\ref{figure_drop_plots}e, with the error bars omitted for clarity.
In Fig.~\ref{figure_drop_plots}a, we see that computer science, non-STEM, and psychology students are the least likely to drop their major, while physics and mathematics students are the most likely to drop.
We note that the relatively low drop rate of computer science majors could be due to the late declaration of the computer science major seen in Fig.~\ref{figure_peak_avg_term}.
That is, attrition from computer science prior to when students are allowed to declare the major is not accounted for in Fig.~\ref{figure_drop_plots}.

Though the patterns in each subset of students largely mimic the pattern overall in Fig.~\ref{figure_drop_plots}a, there are a few potential exceptions.
Figure~\ref{figure_drop_plots}e draws attention to two majors that may have differential rates of attrition by race/ethnicity (namely, in physics and economics), though the size of the error bars in Figs.~\ref{figure_drop_plots}b, \ref{figure_drop_plots}c, and \ref{figure_drop_plots}d indicates that these differences may not be statistically significant.
In both cases, URM students have the highest rate of attrition, followed by Asian students, and finally White students have the lowest rates of attrition.

In economics, the drop rates range from 34\% of URM economics majors dropping the major (Fig.~\ref{figure_drop_plots}c) to 22\% of White economics majors (Fig.~\ref{figure_drop_plots}d).
These differences could prove to be meaningful as more data becomes available, but the large standard error prohibit strong conclusions from this data.
Physics, despite having even larger standard error due to a small sample size, shows an even larger disparity between URM and White students.
In particular, 65\% of URM physics majors (Fig.~\ref{figure_drop_plots}c), 45\% of Asian physics majors (Fig.~\ref{figure_drop_plots}b), and 34\% of White physics majors (Fig.~\ref{figure_drop_plots}d) drop the major.
While distinguishing between the drop rates of Asian students and other students is not possible given the standard errors, the particularly high drop rate of URM physics students relative to White students may be more meaningful.

\subsection{Trajectories of Students After Dropping a Major}
\label{sec:trajectories}

\begin{figure*}
    \centering
    
	\includegraphics[width=0.95\textwidth]{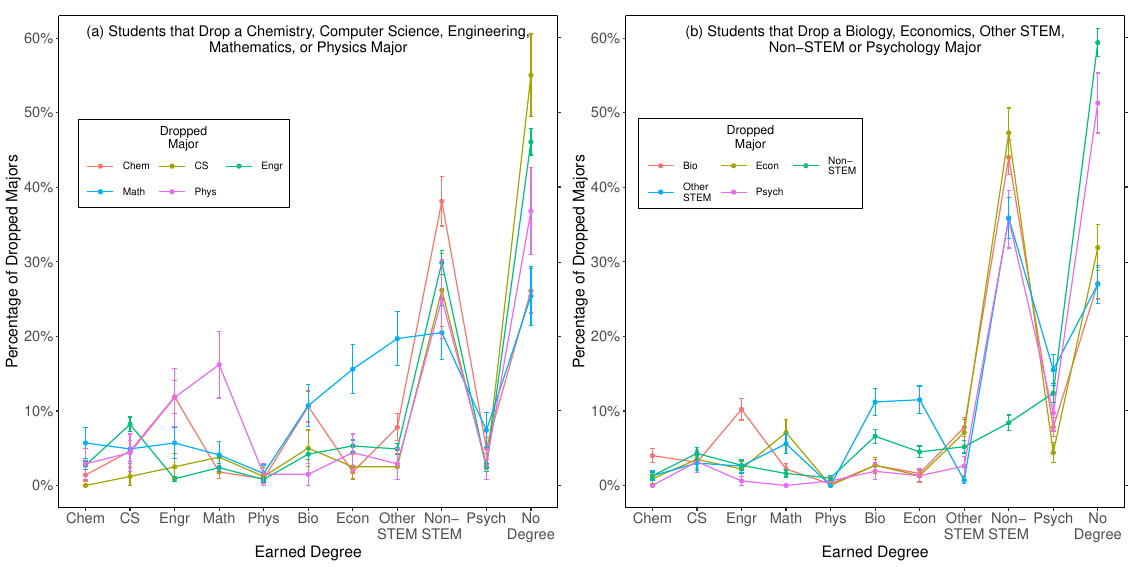}
	
	\caption{\label{figure_drop_plots_degree}
		Among the students that drop each STEM major as well as psychology and non-STEM majors (indicated by line color in the legend), the fractions of students that go on to earn a degree in other majors, or who do not earn a degree at all, are plotted along with their standard error.
		Dropped majors are grouped into (a) chemistry, computer science, engineering, mathematics, and physics and astronomy majors, and (b) biological science, economics, other STEM, psychology, and non-STEM majors.
		}
\end{figure*}

Knowing now how many students drop each major, we answer \ref{rq_droppers} by plotting where those dropped majors ended up in Fig.~\ref{figure_drop_plots_degree}.
In particular, the major indicated in the legends of Fig.~\ref{figure_drop_plots_degree}a and \ref{figure_drop_plots_degree}b shows which major was dropped, while the plot shows which percentage of those who dropped that major ultimately earned a degree in each of the majors on the horizontal axis, including ``no degree.''
For example in Fig.~\ref{figure_drop_plots_degree}a, we see that among the students that drop a physics major (indicated by the line color in the legend), roughly 15\% of them end up earning a degree in mathematics (by looking at this line's value above ``Math'' on the horizontal axis).
We see that the two most common destinations for those who drop any major is either no degree or a degree in non-STEM, except for non-STEM majors who are most likely to earn no degree or a degree in psychology.

Apart from that main signal of dropped STEM majors later earning degrees in non-STEM or leaving the university without a degree, we see a few other interesting spikes.
For instance, those who drop a physics major are likely to earn a degree in mathematics (Fig.~\ref{figure_drop_plots_degree}a) and those who drop chemistry and physics (Fig.~\ref{figure_drop_plots_degree}a) as well as biological science (Fig.~\ref{figure_drop_plots_degree}b) are likely to earn engineering degrees.
Further, those who drop from the ``other STEM'' category (geology and environmental science, neuroscience, and statistics) are likely to major in economics and biology (Fig.~\ref{figure_drop_plots_degree}b).
While all students who drop any major are very likely to earn no degree, the percentage of dropped majors in this category exceeds 50\% for computer science (Fig.~\ref{figure_drop_plots_degree}a), non-STEM, and psychology.

\begin{figure*}
    \centering
    
	\includegraphics[width=0.95\textwidth]{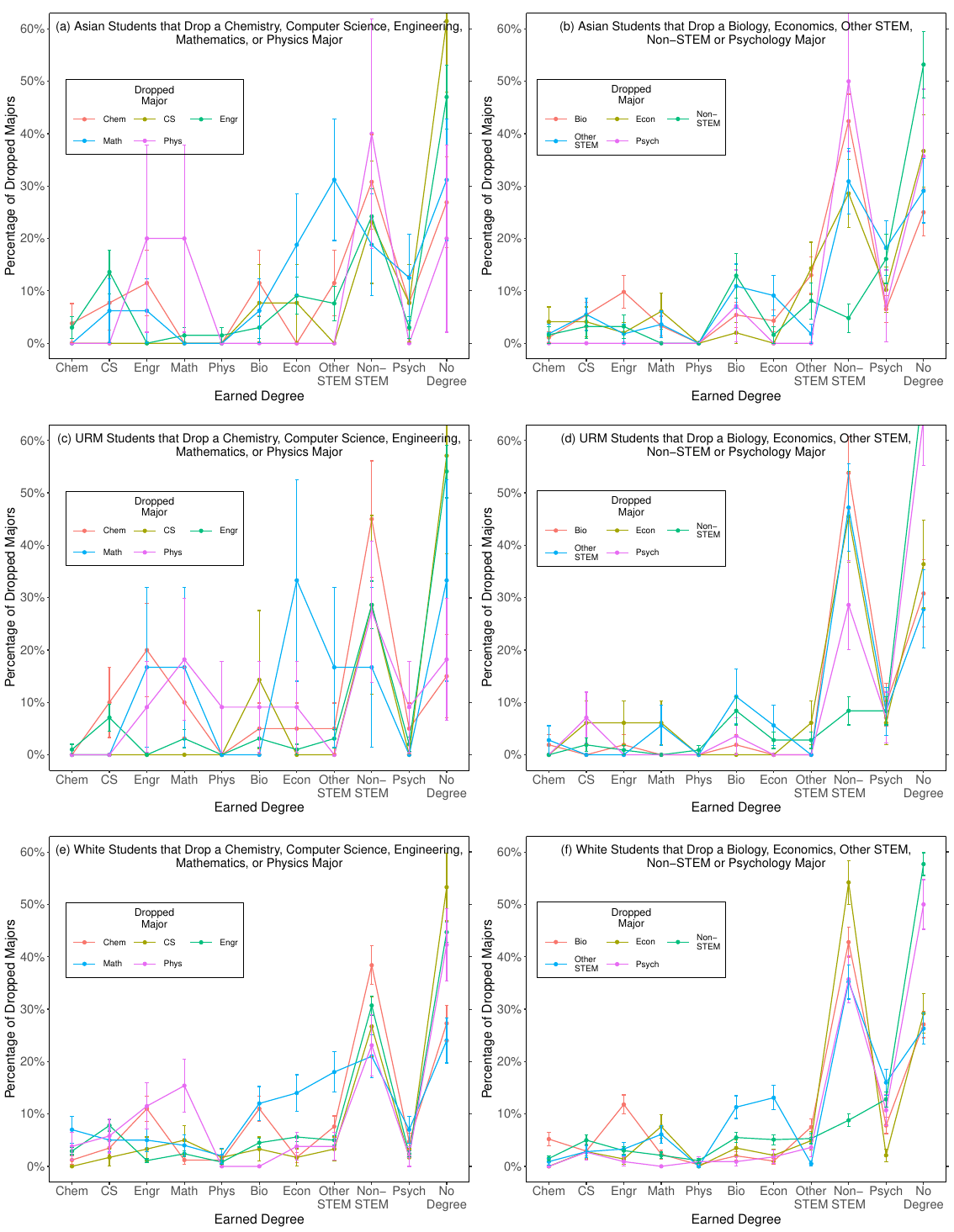}
	
	\caption{\label{figure_drop_plots_degree_by_group}
		Among the students that drop each STEM major as well as psychology (indicated by line color in the legend), the percentage of Asian, URM, and White students that go on to earn a degree in the various other majors, or who do not earn a degree at all, are plotted along with their standard error.
		Dropped majors are grouped into chemistry, computer science, engineering, mathematics, and physics and astronomy majors who are (a) Asian, (c) URM, and (e) White students, and biological science, economics, other STEM, and psychology majors who are (b) Asian, (d) URM, and (f) White students.
		}
\end{figure*}

In order to further answer \ref{rq_droppers}, Fig.~\ref{figure_drop_plots_degree_by_group} plots these same proportions of degrees earned by students who drop a major separately for Asian students (Figs.~\ref{figure_drop_plots_degree_by_group}a and \ref{figure_drop_plots_degree_by_group}b), URM students (Figs.~\ref{figure_drop_plots_degree_by_group}c and \ref{figure_drop_plots_degree_by_group}d), and White students (Figs.~\ref{figure_drop_plots_degree_by_group}e and \ref{figure_drop_plots_degree_by_group}f).
We see for the most part very similar patterns between Asian, URM, and White students, with a few notable differences.
For example, among students who drop a biology major, we see that roughly 10-12\% of the Asian students (Fig.~\ref{figure_drop_plots_degree_by_group}b) and White students (Fig.~\ref{figure_drop_plots_degree_by_group}f)  go on the earn a degree in engineering, compared with roughly 2\% of URM students (Fig.~\ref{figure_drop_plots_degree_by_group}d).
We see a similar pattern among students who drop a chemistry major, with roughly 10\% of the Asian and White students subsequently earning a degree in biology (Figs.~\ref{figure_drop_plots_degree_by_group}a and \ref{figure_drop_plots_degree_by_group}e) compared with roughly 5\% of URM students (Fig.~\ref{figure_drop_plots_degree_by_group}c).

A more detailed accounting of the degrees earned by students who drop each major is provided in Tables~\ref{table_appendix_all_supp}, \ref{table_appendix_asian_supp}, \ref{table_appendix_urm_supp}, and \ref{table_appendix_white_supp} in Appendix C.

\subsection{Degree-Earning Rates}
\label{sec:degree_earning}

\begin{figure*}
    \centering
    
	\includegraphics[width=0.95\linewidth]{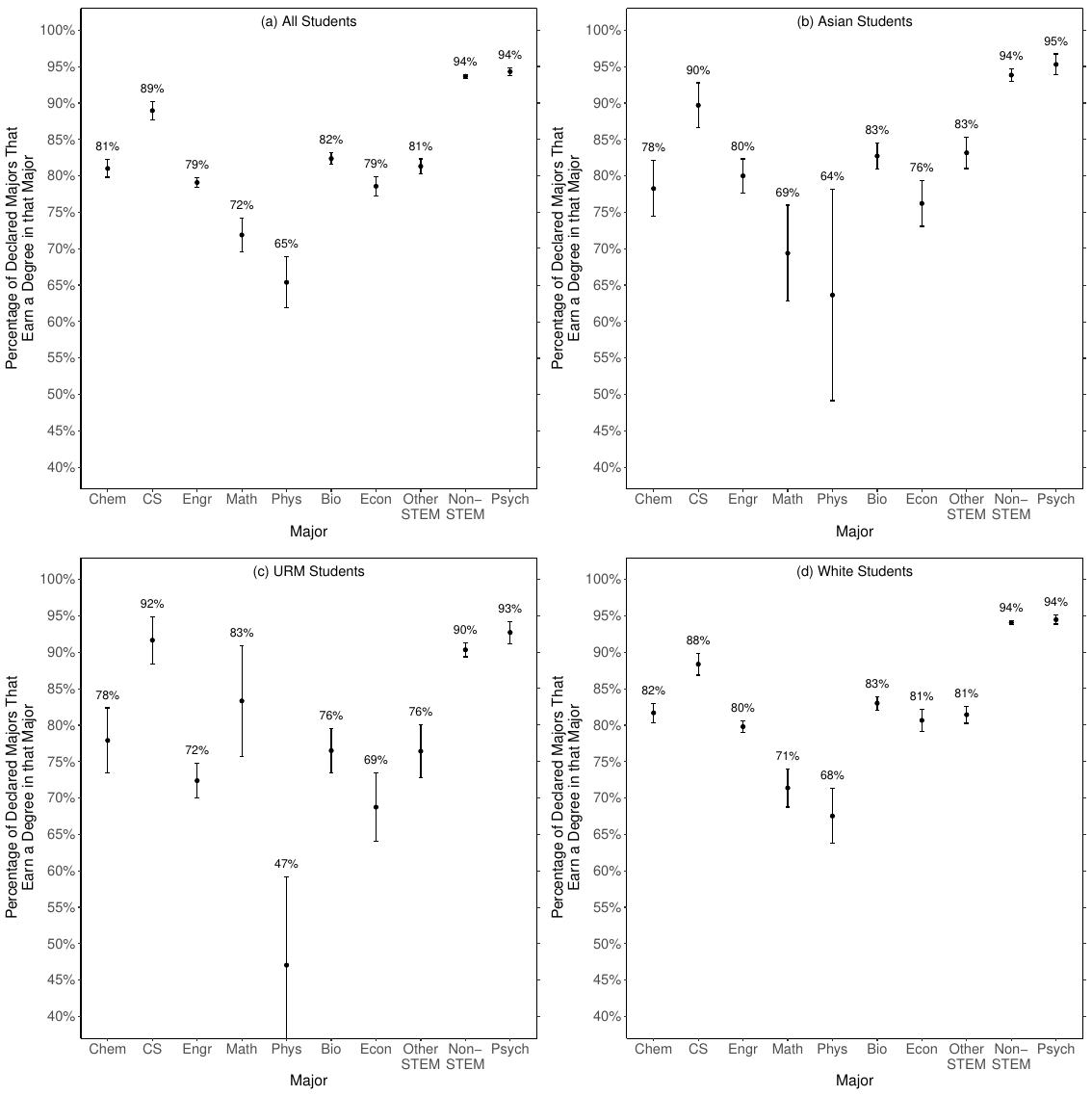}
	
	\caption{\label{figure_degree_rates}
		For each major listed on the horizontal axis, the percentages of (a) all students, (b) Asian students, (c) URM students, and (d) White students who declare that major and then earn a degree in that major are plotted along with the standard error.
		}
\end{figure*}

In order to answer \ref{rq_degree}, we investigated how many students successfully earn a degree in each major.
Figure~\ref{figure_degree_rates}a shows these degree-earning rates for all students in each major, while Fig.~\ref{figure_degree_rates}b shows these rates for Asian students, Fig.~\ref{figure_degree_rates}c for URM students, and Fig.~\ref{figure_degree_rates}d for White students.
While these are broadly similar to an inverse of the drop rates in Fig.~\ref{figure_drop_plots}, since some students drop a major and subsequently declare the same major again, these degree-earning rates are a more direct measurement of persistence in a major.

Looking first at the overall rates in Fig.~\ref{figure_degree_rates}a, there are fairly wide differences across majors, from the lowest rate in physics of about 65\% to the highest in psychology and non-STEM, each at about 94\%.
The highest degree-earning rate in STEM occurs in computer science, with about 89\% of declared computer science majors completing the degree requirements.
As in Fig.~\ref{figure_drop_plots}, this can be at least partially explained by the requirements prior to declaring the major, which causes only students who have already progressed through a significant portion of the computer science curriculum to have declared a computer science major.

Considering then the differences among the different racial/ethnic groups, we see relatively few differences in these degree-earning rates.
As in Fig.~\ref{figure_drop_plots}, the largest difference seen here appears to be in physics, with 68\% of White physics majors earning a physics degree (Fig.~\ref{figure_degree_rates}d) compared to only 47\% of URM physics majors (Fig.~\ref{figure_degree_rates}c), with Asian students very similar to White students with 64\% earning a physics degree (Fig.~\ref{figure_degree_rates}b), albeit with large standard error driven by the low sample size in physics shown in Fig.~\ref{figure_n_unique}.
Similarly, White (Fig.~\ref{figure_degree_rates}d) and Asian students (Fig.~\ref{figure_degree_rates}b) are more likely to complete a degree in economics than URM students (Fig.~\ref{figure_degree_rates}c), but again the size of the standard error prevents any conclusive statements about this difference.

Across all of Fig.~\ref{figure_degree_rates}, we note that since we have combined many majors for the ``non-STEM'' category, this is only a measure of the number of non-STEM majors who successfully earn a degree in any non-STEM major.
That is, a student who drops one non-STEM major but earns a degree in a different non-STEM major will still be counted as having successfully earned a non-STEM degree.
The same is true for the ``other STEM'' and ``engineering'' categories which also combine several majors.
The high ``success rates'' of computer science and psychology may be due in part to the structure of their program encouraging students to declare slightly later than other disciplines, and so this measure may not be capturing attrition that happens prior to an official declaration of major (e.g., a student intending to major in a discipline decides against it before ever declaring that major).
On the other hand, since all students enrolled in the engineering school are considered ``undeclared engineering,'' major, the relatively low degree-earning rate of engineering reflects attrition even from the first to the second term, which is not captured for many other majors in which most students have not yet formally declared a major in their first term.
Thus, each reported degree-earning rate here is a ceiling on the true rate that would include those students who intend to major but never declare.

\subsection{Mean GPA of Degree-Earners vs. Major-Droppers}
\label{sec:gpa}

\begin{figure*}
    \centering
    
	\includegraphics[width=0.95\textwidth]{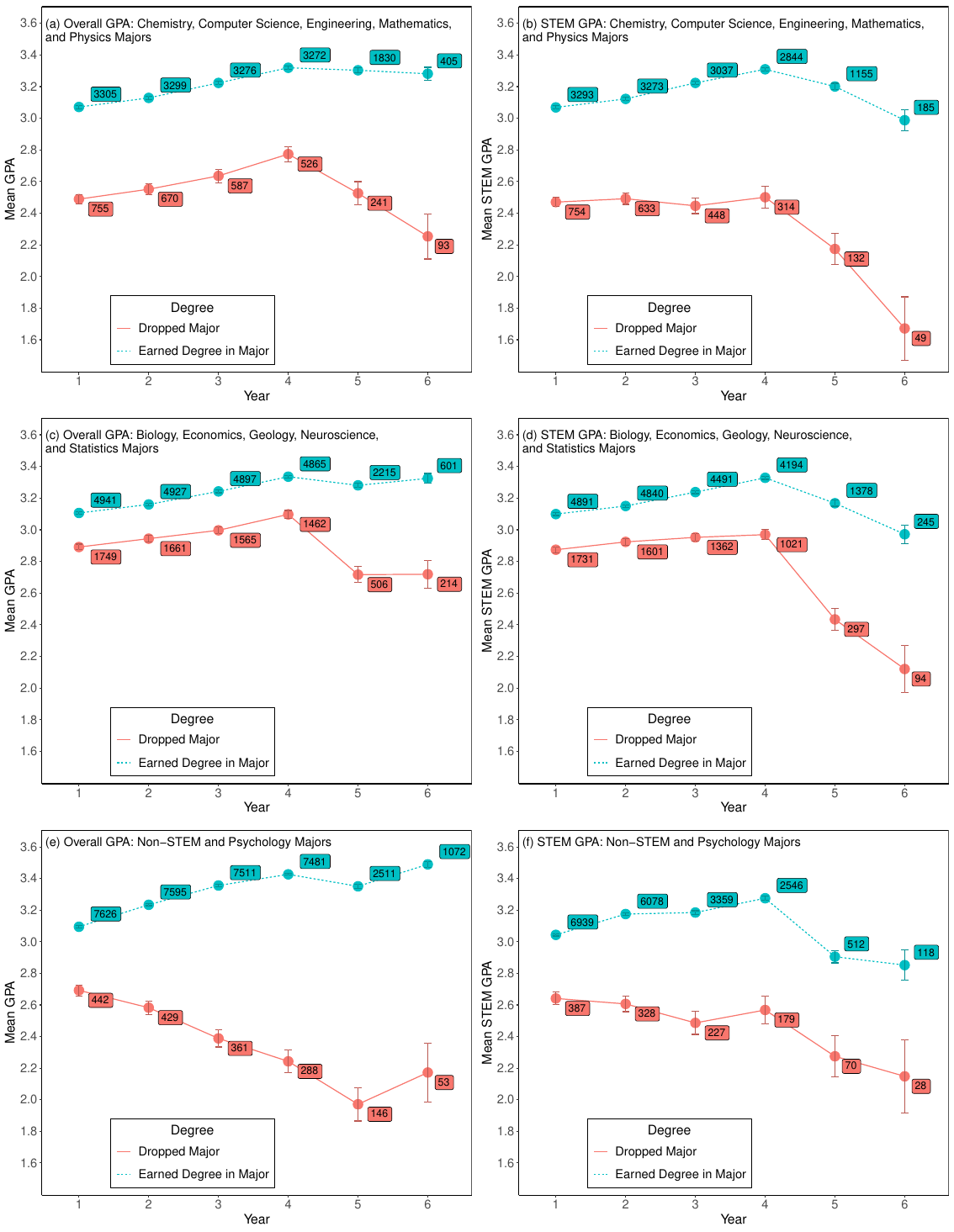}
	
	\caption{\label{figure_gpa_major}
		GPA and STEM GPA over time.
		Each GPA is calculated yearly, not cumulatively.
		Majors are divided into three groupings: (a) and (b) chemistry, computer science, engineering, mathematics, and physics; (c) and (d) biology, economics, geology, neuroscience, and statistics; and (e) and (f) non-STEM including psychology.
		GPA in all courses -- (a), (c), and (e) -- and in only STEM courses -- (b), (d), and (f) -- are calculated separately for two categories of students that declared at least one of the majors in each group: those that ultimately earned a degree in that group of majors and those that dropped from that group of majors.
		For each group, the mean GPA is plotted along with its standard error, with the sample size listed above each point and guides to the eye connecting the points.
		}
\end{figure*}

In order to further our understanding of why students may have dropped a given major and answer \ref{rq_gpa}, Fig.~\ref{figure_gpa_major} plots the mean GPA of students who declared different sets of majors and then either earned a degree within that set of majors or dropped those majors.
Note that students who dropped a major could have gone on to earn a degree with a different major or left the university without a degree.
Both overall GPA (Figs.~\ref{figure_gpa_major}a, \ref{figure_gpa_major}c, and \ref{figure_gpa_major}e) and STEM GPA (Figs.~\ref{figure_gpa_major}b, \ref{figure_gpa_major}d, and \ref{figure_gpa_major}f) are plotted.
Across all of Fig.~\ref{figure_gpa_major}, the large drop in sample size from year four to five and again from five to six is primarily due to students graduating.

We observe that in general, the students who drop that major have a lower GPA and STEM GPA than students who earned a degree in that major.
However, the difference between the two groups varies based on which cluster of majors we consider.
For chemistry, computer science, engineering, mathematics, and physics and astronomy majors (Figs.~\ref{figure_gpa_major}a and \ref{figure_gpa_major}b), those that earned a degree in this set of majors have a GPA of roughly 0.6 grade points higher than those that dropped.
This is consistent through the first four years of study, and similar between overall GPA and STEM GPA.
This difference in grade points represents a difference of one to two letter grades at the studied university, where, for example, the difference between a B and B$+$ is 0.25 grade points and between a B$+$ and A$-$ is 0.5 grade points.
Further, the number of students dropping from this set of majors is roughly 19\% of the total.

For biological science, economics, geology and environmental science, neuroscience, and statistics majors (Fig.~\ref{figure_gpa_major}c and \ref{figure_gpa_major}d), those that earned a degree in this set of majors have a GPA of roughly 0.2 grade points higher than those that dropped, with roughly 26\% of majors dropping.
As with the first set of majors, this is consistent between overall GPA and STEM GPA, and across the first four years of study.

Finally, for non-STEM majors including psychology (Fig.~\ref{figure_gpa_major}e and \ref{figure_gpa_major}f), the overall GPA disparity widens over time from roughly 0.4 grade points in the first year to roughly 1.2 grade points in the fourth year, while in STEM courses the GPA disparity rises from roughly 0.4 grade points in the first year to roughly 0.7 grade points in the fourth year.
Notably, a much smaller fraction of students are dropping from this set -- about 5\% of the total number of students -- which could be due in part to the wide net of considering all non-STEM majors.

\begin{figure*}
    \centering
    
	\includegraphics[width=0.95\textwidth]{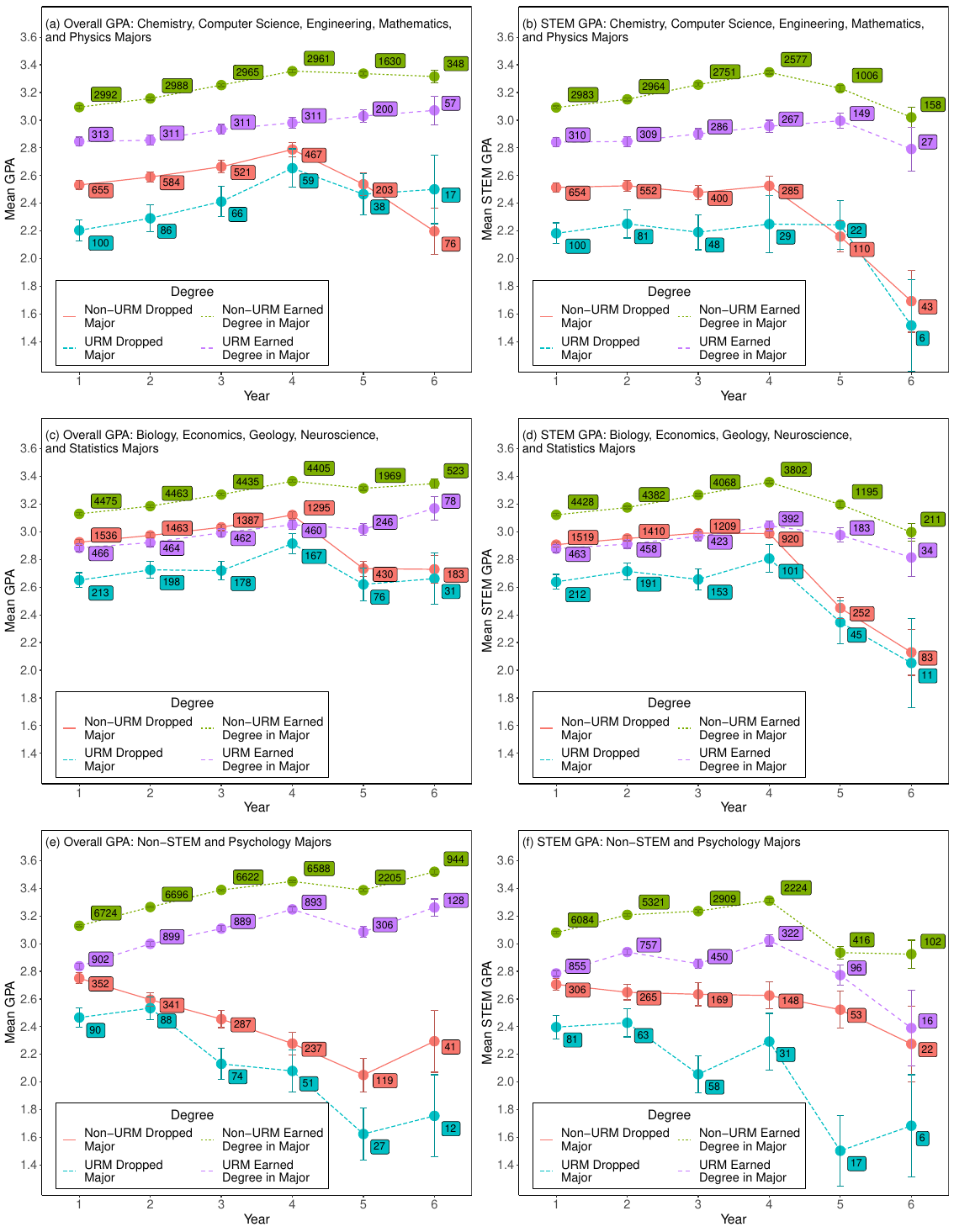}
	
	\caption{\label{figure_race_gpa_major}
		GPA and STEM GPA over time by racial/ethnic group.
		Each GPA is calculated yearly, not cumulatively.
		Majors are divided into three groupings: (a) and (b) chemistry, computer science, engineering, mathematics, and physics; (c) and (d) biology, economics, geology, neuroscience, and statistics; and (e) and (f) non-STEM including psychology.
		GPA in all courses -- (a), (c), and (e) -- and in only STEM courses -- (b), (d), and (f) -- are calculated separately for four categories of students that declared at least one of the majors in each group: URM and non-URM (i.e., White and Asian) students that ultimately earned a degree in that group of majors and those that dropped from that group of majors.
		For each group, the mean GPA is plotted along with its standard error, with the sample size listed above each point and guides to the eye connecting the points.
		}
\end{figure*}

We further consider the same measures separately for URM and non-URM (i.e., Asian and White) students in Fig.~\ref{figure_race_gpa_major}.
In this case, Asian and White students have been combined to improve clarity in the figure since there were few differences between these two groups when considering them separately.
A version of Fig.~\ref{figure_race_gpa_major} that separately plots Asian and White students can be seen in Fig.~\ref{figure_race_gpa_major_all} in Appendix D.

Figure~\ref{figure_race_gpa_major} has the same subfigure structure as Fig.~\ref{figure_gpa_major}.
That is, overall GPA is plotted in Figs.~\ref{figure_race_gpa_major}a, \ref{figure_race_gpa_major}c, and \ref{figure_race_gpa_major}e and STEM GPA in Figs.~\ref{figure_race_gpa_major}b, \ref{figure_race_gpa_major}d, and \ref{figure_race_gpa_major}f.
And students belonging to different clusters of majors are considered separately: chemistry, computer science, engineering, mathematics, and physics and astronomy majors are included in Figs.~\ref{figure_race_gpa_major}a and \ref{figure_race_gpa_major}b; biological science, economics, geology and environmental science, neuroscience, and statistics majors in Figs.~\ref{figure_race_gpa_major}c and \ref{figure_race_gpa_major}d; and non-STEM and psychology majors in Figs.~\ref{figure_race_gpa_major}e and \ref{figure_race_gpa_major}f.

In all cases, the main finding is that URM students are earning lower grades on average than non-URM students, both among those who drop a given major and those who earn a degree in that major.
The degree of this disparity differs by which majors we consider, but among STEM majors it appears relatively similar whether we consider overall GPA (Figs.~\ref{figure_race_gpa_major}a and \ref{figure_race_gpa_major}c) or STEM GPA (Figs.~\ref{figure_race_gpa_major}b and \ref{figure_race_gpa_major}d).
In particular, among chemistry, computer science, engineering, mathematics, and physics and astronomy majors, URM students who earn a degree have a GPA (Fig.~\ref{figure_race_gpa_major}a) and STEM GPA (Fig.~\ref{figure_race_gpa_major}b) roughly 0.3 grade points lower on average than non-URM students who earn a degree.
Among the same majors but considering now students who drop the major, URM students again have roughly 0.3 grade points lower than non-URM students in overall (Fig.~\ref{figure_race_gpa_major}a) and STEM GPA (Fig.~\ref{figure_race_gpa_major}b), with the gap in overall GPA (Fig.~\ref{figure_race_gpa_major}a) somewhat closing by the fourth year.

There is a similar trend among biological sciences, economics, geology, neuroscience, and statistics majors (Figs.~\ref{figure_race_gpa_major}c and \ref{figure_race_gpa_major}d), with URM students having roughly 0.2-0.3 grade points lower than non-URM students, among those students who earn a degree in those majors and those who drop these majors.
Similarly among non-STEM majors including psychology, there is an 0.3-0.4 grade point difference between URM and non-URM students in overall GPA (Fig.~\ref{figure_race_gpa_major}e) and STEM GPA (Fig.~\ref{figure_race_gpa_major}f).

\section{Discussion}
\label{discussion}

In this section, we first discuss general trends and then follow up with a discussion of race/ethnicity differences.

\subsection{General Enrollment Patterns}
\label{sec:discuss_general_enrollment}

Despite large differences in the number of students enrolling in different STEM disciplines at the studied university (Fig.~\ref{figure_n_unique}a), there are broadly similar patterns of when those students declare the major (Fig.~\ref{figure_peak_avg_term}a), with some exceptions (i.e., engineering and computer science due to the constraints on when a student can declare a major).
However, there are notable differences in the attrition of students from the different majors (Fig.~\ref{figure_drop_plots}a), and the corresponding degree-earning rates (Fig.~\ref{figure_degree_rates}a).
Notably, a few STEM disciplines stand out as having particularly high rates of attrition (or low rates of degree completion for students who declared those majors),e.g., mathematics and physics.
This is consistent with a study by Leslie \textit{et al.}  which identifies mathematics and physics as the two STEM disciplines with the highest ``ability belief'' (i.e., emphasis on brilliance)~\cite{leslie2015}.
This trend of high attrition rates is particularly problematic for mathematics and physics, since these disciplines recruit very few students in the first place (Fig.~\ref{figure_n_unique}).
Moreover, mathematics and physics are also two disciplines with deeply hierarchical knowledge structures, which could influence student decision making, e.g., whether to leave the discipline after unsatisfactory experiences in earlier courses.

\subsection{Enrollment Patterns by Race/Ethnicity}
\label{sec:discuss_race_enrollment}

The most notable example of race/ethnicity differences in enrollment patterns observed in our analysis is in Fig.~\ref{figure_percentages}.
We observe that in biological science, geological and environmental science, neuroscience, and economics, Asian students enroll at a higher rate than URM or White students.
This trend is hinted at in Fig.~\ref{figure_percentages}a, but made clear in Fig.~\ref{figure_percentages}b where we see that a substantially higher percentage of Asian students choose these majors relative to URM and White students.
Correspondingly, a lower percentage of Asian students choose non-STEM majors than URM and White students, and similarly fewer Asian students choose engineering majors.
There are also small differences in just a few other disciplines, such as more URM students majoring in psychology and more White students majoring in engineering relative to their peers.

Apart from these differences, enrollment in the majority of majors is balanced compared to the population as a whole.
This is good since it means that race/ethnicity are not playing a large role in the students' choices of major.
However, it is not an equitable situation since URM students are still not being properly represented in classrooms.
In particular, for URM students, the inequities often present in opportunities for college preparation and admissions are still present in the classrooms, but they are not worsened by the students' choices of major.

Moreover, we see no notable differences by race/ethnicity in when the students declare these majors (Fig.~\ref{figure_peak_avg_term}).
Though the term in which the peak number of students add the major may shift, the average term of adding the major is consistent among Asian, URM, and White students.

\subsection{Attrition and Degree-Earning Rates}
\label{sec:discuss_attrition}

The primary difference in enrollment patterns found is in the attrition rates of students (Fig.~\ref{figure_drop_plots}) and correspondingly the degree-earning rates (Fig.~\ref{figure_degree_rates}).
Though most majors have relatively similar drop rates and degree-earning rates among Asian, URM, and White students, a very notable exception is physics.
Figure~\ref{figure_drop_plots}c shows that 65\% of URM students who declare a physics major ultimately drop that major, which does not even include those students who decided to change their path before officially declaring a physics major.
This is compared to 45\% of Asian physics majors (\ref{figure_drop_plots}b) and 34\% of White physics majors (\ref{figure_drop_plots}d) dropping the major.
This is especially problematic for physics, which recruits very few students to begin with (Fig.~\ref{figure_n_unique}), especially among Asian (Fig.~\ref{figure_n_unique}b) and URM students (Fig.~\ref{figure_n_unique}c).
The fact that almost two-thirds of URM students drop the physics major compared to only roughly one-third of White students may be a sign of a lack of appropriate support for students and an inequitable learning environment.
Further, this is consistent with prior research showing that underserved students in the sciences are especially vulnerable to stereotype threat~\cite{astin1993, cross1993, bianchini2002, britner2006, bianchini2013, basile2015, hilts2018}, especially in physics which is perceived as a field that requires a high innate ``brilliance'' to succeed~\cite{leslie2015}.

Though no other disciplines show the same degree of racial/ethnic differences in attrition rates, Fig.~\ref{figure_drop_plots}e shows that with the exception of computer science and mathematics, URM students have slightly higher attrition rates from every major.
This is a deeply troubling trend, since it runs counter to the somewhat hopeful situation noted in the balanced declarations of major (at least with regard to race/ethnicity) seen in Fig.~\ref{figure_percentages}.
Thus, although they initially have similar aspirations as their overrepresented peers, URM students in general, including both STEM and non-STEM discplines, are being forced to change their academic plans at higher rates than Asian and White students.

Though we had hoped to be able to shed some more light on what happens to the students who drop each major, the low sample size in Fig.~\ref{figure_drop_plots_degree_by_group} prevents any notable conclusions.
This aspect of student attrition from various majors can be revisited as more data becomes available. 

\subsection{Racial/Ethnic GPA Differences}
\label{sec:discuss_gpa}

We find pervasive and troubling trends in the overall GPA and STEM GPA of STEM majors.
We find in Figs.~\ref{figure_race_gpa_major}a and \ref{figure_race_gpa_major}b that among students who major in chemistry, computer science, engineering, mathematics, and physics, URM students consistently earn lower grades than their non-URM peers.
This is true both among students that persist in their degree attainment and those that drop the major.
This is problematic since these URM students are already forced to contend with stereotypes that pressure them away from STEM, and given the intricate relationship between students' expectancy or self-efficacy and academic performance~\cite{britner2006, zimmerman2000, pintrich2003, bandura1991, bandura1994, bandura1997, bandura1999, bandura2001, bandura2005}, the feedback that these students obtain only compounds the existing societal pressures working against them and widens the racial/ethnic gap further.

Moreover, among students in the remaining STEM majors (biological sciences, economics, geology and environmental science, neuroscience, and statistics; Figs.~\ref{figure_race_gpa_major}c and \ref{figure_race_gpa_major}d), the URM students who persist and earn a degree earn on average the same grades as the Asian and White students who drop the major. The reason for these troubling trends must be investigated further because they may signify lack of sufficient support, mentoring, and guidance to ensure excellence of the URM students who are already severely disadvantaged.

Students must constantly make decisions about their academic future, and Expectancy Value Theory states that students' performance
 is influenced by societal stereotypes, made worse by a non-inclusive classroom learning environment in which they are underrepresented as well as previous performance feedback received.
Many of these students, who come into college with a somewhat weaker preparation than their peers through no fault of their own but due to historical disadvantages, must then face these extreme pressures against their success in various STEM fields.
It is critical that universities make concerted efforts to mentor and support all of their students, and improve the learning environment of these STEM disciplines to counteract the historically-rooted culture and stereotypes surrounding STEM that unfairly disadvantage URM students.

\subsection{Limitations and Implications}
\label{sec:discuss_limitations}
One limitation of this study is the fields with the most consistent racial/ethnic differences, particularly physics, also have the lowest number of students which limits statistical power.
Also, this study limits its considerations to race/ethnicity.
Other studies with larger data sets could investigate how other underserved populations such as women, first-generation college students, or low-income students may be disadvantaged in STEM.

A critically important extension of this work would be for other institutions of different types and sizes to do similar analyses in order to broaden the wealth of knowledge available and continue to work towards the goal of equitable and inclusive education.
Other institutions noting similar problematic trends can help pinpoint common sources of inequities, while institutions that do not observe these trends may be able to identify how they have structured their programs to avoid these inequitable trends.
Studies such as this can thus provide a framework for other institutions to perform similar analyses, and for particular departments to understand how their own trends differ from those of other departments at their own university.

Focus on increasing equity and inclusion in learning is especially important in the early courses for these STEM majors, since they are fraught with problematic racial/ethnic differences and may be a significant source of URM students dropping the majors.
All STEM instructors should consider how they are supporting their students throughout the curriculum, since even among those that did not show differences in attrition rates we still see problematic GPA differences between URM and non-URM students.
All of these issues should be addressed since they are critical for improving equity and inclusion in higher education STEM learning environments.

\begin{acknowledgements}
This research is supported by the National Science Foundation Grant DUE-1524575 and the Sloan Foundation Grant G-2018-11183.
\end{acknowledgements}

\bibliography{refs}   

\clearpage
\onecolumngrid
\appendix
\section{Appendix A: Number of Current Majors and those who Added or Dropped each Major by Term}
\label{appendix_A}

\begin{table*}[b]

	\centering
	
	\begin{minipage}{0.45\linewidth}%
		\begin{center}
	(a) \textbf{Chemistry}, $N_{\textrm{unique}} = 1063$
\end{center}

\begin{tabular}{c | r r r}
	\multirow{2}{*}{Term}	& \multicolumn{3}{c}{Number of Majors [\% of $N_{\textrm{unique}}$]} \\
	& Current	& Added	& Dropped	\\
	\hline
1	& 13 \hphantom{1}[1.2]	& 13 \hphantom{1}[1.2]	& 1 \hphantom{1}[0.1]	\\
2	& 99 \hphantom{1}[9.3]	& 86 \hphantom{1}[8.1]	& 1 \hphantom{1}[0.1]	\\
3	& 800 [75.3]	& 706 [66.4]	& 10 \hphantom{1}[0.9]	\\
4	& 890 [83.7]	& 147 [13.8]	& 59 \hphantom{1}[5.6]	\\
5	& 919 [86.5]	& 70 \hphantom{1}[6.6]	& 42 \hphantom{1}[4.0]	\\
6	& 904 [85.0]	& 23 \hphantom{1}[2.2]	& 39 \hphantom{1}[3.7]	\\
7	& 887 [83.4]	& 13 \hphantom{1}[1.2]	& 25 \hphantom{1}[2.4]	\\
8	& 861 [81.0]	& 1 \hphantom{1}[0.1]	& 15 \hphantom{1}[1.4]	\\
9	& 403 [37.9]	& 6 \hphantom{1}[0.6]	& 10 \hphantom{1}[0.9]	\\
10	& 157 [14.8]	& 1 \hphantom{1}[0.1]	& 7 \hphantom{1}[0.7]	\\
11	& 30 \hphantom{1}[2.8]	& 0 \hphantom{1}[0.0]	& 5 \hphantom{1}[0.5]	\\
12	& 14 \hphantom{1}[1.3]	& 1 \hphantom{1}[0.1]	& 4 \hphantom{1}[0.4]	\\
\end{tabular}
		
		\vspace{3mm}
		
		\begin{center}
	(c) \textbf{Engineering}, $N_{\textrm{unique}} = 3505$
\end{center}

\begin{tabular}{c | r r r}
	\multirow{2}{*}{Term}	& \multicolumn{3}{c}{Number of Majors [\% of $N_{\textrm{unique}}$]} \\
	& Current	& Added	& Dropped	\\
	\hline
1	& 3164 [90.3]	& 3164 [90.3]	& 46 \hphantom{1}[1.3]	\\
2	& 2919 [83.3]	& 14 \hphantom{1}[0.4]	& 259 \hphantom{1}[7.4]	\\
3	& 2908 [83.0]	& 158 \hphantom{1}[4.5]	& 169 \hphantom{1}[4.8]	\\
4	& 2841 [81.1]	& 53 \hphantom{1}[1.5]	& 120 \hphantom{1}[3.4]	\\
5	& 2841 [81.1]	& 62 \hphantom{1}[1.8]	& 62 \hphantom{1}[1.8]	\\
6	& 2841 [81.1]	& 27 \hphantom{1}[0.8]	& 27 \hphantom{1}[0.8]	\\
7	& 2822 [80.5]	& 20 \hphantom{1}[0.6]	& 38 \hphantom{1}[1.1]	\\
8	& 2788 [79.5]	& 5 \hphantom{1}[0.1]	& 25 \hphantom{1}[0.7]	\\
9	& 1586 [45.2]	& 4 \hphantom{1}[0.1]	& 15 \hphantom{1}[0.4]	\\
10	& 627 [17.9]	& 1 \hphantom{1}[0.0]	& 15 \hphantom{1}[0.4]	\\
11	& 112 \hphantom{1}[3.2]	& 0 \hphantom{1}[0.0]	& 7 \hphantom{1}[0.2]	\\
12	& 48 \hphantom{1}[1.4]	& 0 \hphantom{1}[0.0]	& 7 \hphantom{1}[0.2]	\\
\end{tabular}
	\end{minipage}\hfill%
	\begin{minipage}{0.45\linewidth}%
		\begin{center}
	(b) \textbf{Computer Science}, $N_{\textrm{unique}} = 603$
\end{center}

\begin{tabular}{c | r r r}
	\multirow{2}{*}{Term}	& \multicolumn{3}{c}{Number of Majors [\% of $N_{\textrm{unique}}$]} \\
	& Current	& Added	& Dropped	\\
	\hline
1	& 1 \hphantom{1}[0.2]	& 1 \hphantom{1}[0.2]	& 0 \hphantom{1}[0.0]	\\
2	& 20 \hphantom{1}[3.3]	& 19 \hphantom{1}[3.2]	& 0 \hphantom{1}[0.0]	\\
3	& 97 [16.1]	& 78 [12.9]	& 1 \hphantom{1}[0.2]	\\
4	& 208 [34.5]	& 113 [18.7]	& 5 \hphantom{1}[0.8]	\\
5	& 387 [64.2]	& 184 [30.5]	& 6 \hphantom{1}[1.0]	\\
6	& 450 [74.6]	& 73 [12.1]	& 10 \hphantom{1}[1.7]	\\
7	& 489 [81.1]	& 61 [10.1]	& 17 \hphantom{1}[2.8]	\\
8	& 484 [80.3]	& 26 \hphantom{1}[4.3]	& 10 \hphantom{1}[1.7]	\\
9	& 234 [38.8]	& 36 \hphantom{1}[6.0]	& 10 \hphantom{1}[1.7]	\\
10	& 144 [23.9]	& 11 \hphantom{1}[1.8]	& 10 \hphantom{1}[1.7]	\\
11	& 67 [11.1]	& 1 \hphantom{1}[0.2]	& 9 \hphantom{1}[1.5]	\\
12	& 34 \hphantom{1}[5.6]	& 2 \hphantom{1}[0.3]	& 2 \hphantom{1}[0.3]	\\
\end{tabular}
		
		\vspace{3mm}
		
		\begin{center}
	(d) \textbf{Mathematics}, $N_{\textrm{unique}} = 367$
\end{center}

\begin{tabular}{c | r r r}
	\multirow{2}{*}{Term}	& \multicolumn{3}{c}{Number of Majors [\% of $N_{\textrm{unique}}$]} \\
	& Current	& Added	& Dropped	\\
	\hline
1	& 22 \hphantom{1}[6.0]	& 22 \hphantom{1}[6.0]	& 0 \hphantom{1}[0.0]	\\
2	& 87 [23.7]	& 66 [18.0]	& 1 \hphantom{1}[0.3]	\\
3	& 181 [49.3]	& 102 [27.8]	& 10 \hphantom{1}[2.7]	\\
4	& 254 [69.2]	& 94 [25.6]	& 23 \hphantom{1}[6.3]	\\
5	& 265 [72.2]	& 36 \hphantom{1}[9.8]	& 26 \hphantom{1}[7.1]	\\
6	& 280 [76.3]	& 26 \hphantom{1}[7.1]	& 13 \hphantom{1}[3.5]	\\
7	& 272 [74.1]	& 14 \hphantom{1}[3.8]	& 15 \hphantom{1}[4.1]	\\
8	& 258 [70.3]	& 4 \hphantom{1}[1.1]	& 13 \hphantom{1}[3.5]	\\
9	& 80 [21.8]	& 8 \hphantom{1}[2.2]	& 9 \hphantom{1}[2.5]	\\
10	& 60 [16.3]	& 1 \hphantom{1}[0.3]	& 2 \hphantom{1}[0.5]	\\
11	& 18 \hphantom{1}[4.9]	& 1 \hphantom{1}[0.3]	& 6 \hphantom{1}[1.6]	\\
12	& 7 \hphantom{1}[1.9]	& 0 \hphantom{1}[0.0]	& 4 \hphantom{1}[1.1]	\\
\end{tabular}
	\end{minipage}
	
		\vspace{3mm}
		
		\begin{center}
	(e) \textbf{Physics \& Astronomy}, $N_{\textrm{unique}} = 182$
\end{center}

\begin{tabular}{c | r r r}
	\multirow{2}{*}{Term}	& \multicolumn{3}{c}{Number of Majors [\% of $N_{\textrm{unique}}$]} \\
	& Current	& Added	& Dropped	\\
	\hline
1	& 17 \hphantom{1}[9.3]	& 17 \hphantom{1}[9.3]	& 0 \hphantom{1}[0.0]	\\
2	& 72 [39.6]	& 56 [30.8]	& 1 \hphantom{1}[0.5]	\\
3	& 122 [67.0]	& 62 [34.1]	& 14 \hphantom{1}[7.7]	\\
4	& 148 [81.3]	& 33 [18.1]	& 7 \hphantom{1}[3.8]	\\
5	& 144 [79.1]	& 7 \hphantom{1}[3.8]	& 12 \hphantom{1}[6.6]	\\
6	& 140 [76.9]	& 4 \hphantom{1}[2.2]	& 8 \hphantom{1}[4.4]	\\
7	& 131 [72.0]	& 2 \hphantom{1}[1.1]	& 11 \hphantom{1}[6.0]	\\
8	& 125 [68.7]	& 1 \hphantom{1}[0.5]	& 6 \hphantom{1}[3.3]	\\
9	& 46 [25.3]	& 1 \hphantom{1}[0.5]	& 4 \hphantom{1}[2.2]	\\
10	& 38 [20.9]	& 0 \hphantom{1}[0.0]	& 2 \hphantom{1}[1.1]	\\
11	& 3 \hphantom{1}[1.6]	& 0 \hphantom{1}[0.0]	& 1 \hphantom{1}[0.5]	\\
12	& 1 \hphantom{1}[0.5]	& 0 \hphantom{1}[0.0]	& 2 \hphantom{1}[1.1]	\\
\end{tabular}
	
	\caption{\label{table_adddrop_1}
		For each term from 1 to 12, the current number of declared majors (``Current'') is shown along with the number of majors who newly declared in that term (``Added'') and the number of former majors who dropped the major as of that term (``Dropped'').
		In square brackets next to each measure is the percentage of all unique students who declared that major.
		The five sub-tables show this information for five different majors: (a) chemistry, (b) computer science, (c) engineering, (d) mathematics, and (e) physics and astronomy.
		For example, in (c) we can see that in term 2, there were 2919 students with a declared engineering major, which represents 83.0\% of all students who ever declared engineering.
		Further, 14 students (0.4\% of all engineering students) who did not declare in term 1 added the major in term 2, and 259 students (7.4\% of all engineering students) who were engineering majors in term 1 dropped the major in term 2.
		}
\end{table*}

\begin{table*}[b]
	
	\centering

	\begin{minipage}{0.45\linewidth}%
		\begin{center}
	(a) \textbf{Biological Sciences}, $N_{\textrm{unique}} = 2211$
\end{center}

\begin{tabular}{c | r r r}
	\multirow{2}{*}{Term}	& \multicolumn{3}{c}{Number of Majors [\% of $N_{\textrm{unique}}$]} \\
	& Current	& Added	& Dropped	\\
	\hline
1	& 17 \hphantom{1}[0.8]	& 17 \hphantom{1}[0.8]	& 0 \hphantom{1}[0.0]	\\
2	& 68 \hphantom{1}[3.1]	& 53 \hphantom{1}[2.4]	& 3 \hphantom{1}[0.1]	\\
3	& 1205 [54.5]	& 1141 [51.6]	& 12 \hphantom{1}[0.5]	\\
4	& 1697 [76.8]	& 568 [25.7]	& 87 \hphantom{1}[3.9]	\\
5	& 1933 [87.4]	& 307 [13.9]	& 82 \hphantom{1}[3.7]	\\
6	& 1926 [87.1]	& 71 \hphantom{1}[3.2]	& 84 \hphantom{1}[3.8]	\\
7	& 1878 [84.9]	& 43 \hphantom{1}[1.9]	& 73 \hphantom{1}[3.3]	\\
8	& 1795 [81.2]	& 11 \hphantom{1}[0.5]	& 48 \hphantom{1}[2.2]	\\
9	& 350 [15.8]	& 7 \hphantom{1}[0.3]	& 34 \hphantom{1}[1.5]	\\
10	& 221 [10.0]	& 2 \hphantom{1}[0.1]	& 16 \hphantom{1}[0.7]	\\
11	& 40 \hphantom{1}[1.8]	& 1 \hphantom{1}[0.0]	& 6 \hphantom{1}[0.3]	\\
12	& 13 \hphantom{1}[0.6]	& 0 \hphantom{1}[0.0]	& 5 \hphantom{1}[0.2]	\\
\end{tabular}
		
		\vspace{3mm}
		
		\begin{center}
	(c) \textbf{Psychology}, $N_{\textrm{unique}} = 1767$
\end{center}

\begin{tabular}{c | r r r}
	\multirow{2}{*}{Term}	& \multicolumn{3}{c}{Number of Majors [\% of $N_{\textrm{unique}}$]} \\
	& Current	& Added	& Dropped	\\
	\hline
1	& 6 \hphantom{1}[0.3]	& 6 \hphantom{1}[0.3]	& 0 \hphantom{1}[0.0]	\\
2	& 61 \hphantom{1}[3.5]	& 55 \hphantom{1}[3.1]	& 0 \hphantom{1}[0.0]	\\
3	& 400 [22.6]	& 340 [19.2]	& 4 \hphantom{1}[0.2]	\\
4	& 1001 [56.6]	& 614 [34.7]	& 25 \hphantom{1}[1.4]	\\
5	& 1436 [81.3]	& 458 [25.9]	& 23 \hphantom{1}[1.3]	\\
6	& 1581 [89.5]	& 171 \hphantom{1}[9.7]	& 29 \hphantom{1}[1.6]	\\
7	& 1643 [93.0]	& 94 \hphantom{1}[5.3]	& 13 \hphantom{1}[0.7]	\\
8	& 1536 [86.9]	& 21 \hphantom{1}[1.2]	& 28 \hphantom{1}[1.6]	\\
9	& 262 [14.8]	& 18 \hphantom{1}[1.0]	& 12 \hphantom{1}[0.7]	\\
10	& 134 \hphantom{1}[7.6]	& 3 \hphantom{1}[0.2]	& 10 \hphantom{1}[0.6]	\\
11	& 30 \hphantom{1}[1.7]	& 0 \hphantom{1}[0.0]	& 6 \hphantom{1}[0.3]	\\
12	& 15 \hphantom{1}[0.8]	& 0 \hphantom{1}[0.0]	& 4 \hphantom{1}[0.2]	\\
\end{tabular}
	\end{minipage}\hfill%
	\begin{minipage}{0.45\linewidth}%
		\begin{center}
	(b) \textbf{Economics}, $N_{\textrm{unique}} = 934$
\end{center}

\begin{tabular}{c | r r r}
	\multirow{2}{*}{Term}	& \multicolumn{3}{c}{Number of Majors [\% of $N_{\textrm{unique}}$]} \\
	& Current	& Added	& Dropped	\\
	\hline
1	& 15 \hphantom{1}[1.6]	& 15 \hphantom{1}[1.6]	& 0 \hphantom{1}[0.0]	\\
2	& 98 [10.5]	& 85 \hphantom{1}[9.1]	& 5 \hphantom{1}[0.5]	\\
3	& 290 [31.0]	& 200 [21.4]	& 9 \hphantom{1}[1.0]	\\
4	& 543 [58.1]	& 267 [28.6]	& 21 \hphantom{1}[2.2]	\\
5	& 685 [73.3]	& 170 [18.2]	& 30 \hphantom{1}[3.2]	\\
6	& 740 [79.2]	& 86 \hphantom{1}[9.2]	& 33 \hphantom{1}[3.5]	\\
7	& 742 [79.4]	& 69 \hphantom{1}[7.4]	& 51 \hphantom{1}[5.5]	\\
8	& 694 [74.3]	& 27 \hphantom{1}[2.9]	& 29 \hphantom{1}[3.1]	\\
9	& 179 [19.2]	& 11 \hphantom{1}[1.2]	& 21 \hphantom{1}[2.2]	\\
10	& 89 \hphantom{1}[9.5]	& 4 \hphantom{1}[0.4]	& 15 \hphantom{1}[1.6]	\\
11	& 33 \hphantom{1}[3.5]	& 4 \hphantom{1}[0.4]	& 4 \hphantom{1}[0.4]	\\
12	& 11 \hphantom{1}[1.2]	& 0 \hphantom{1}[0.0]	& 8 \hphantom{1}[0.9]	\\
\end{tabular}
		
		\vspace{3mm}
		
		\begin{center}
	(d) \textbf{Other STEM}, $N_{\textrm{unique}} = 1507$
\end{center}

\begin{tabular}{c | r r r}
	\multirow{2}{*}{Term}	& \multicolumn{3}{c}{Number of Majors [\% of $N_{\textrm{unique}}$]} \\
	& Current	& Added	& Dropped	\\
	\hline
1	& 16 \hphantom{1}[1.1]	& 16 \hphantom{1}[1.1]	& 0 \hphantom{1}[0.0]	\\
2	& 141 \hphantom{1}[9.4]	& 126 \hphantom{1}[8.4]	& 4 \hphantom{1}[0.3]	\\
3	& 762 [50.6]	& 631 [41.9]	& 13 \hphantom{1}[0.9]	\\
4	& 1184 [78.6]	& 466 [30.9]	& 51 \hphantom{1}[3.4]	\\
5	& 1315 [87.3]	& 183 [12.1]	& 55 \hphantom{1}[3.6]	\\
6	& 1303 [86.5]	& 41 \hphantom{1}[2.7]	& 55 \hphantom{1}[3.6]	\\
7	& 1254 [83.2]	& 27 \hphantom{1}[1.8]	& 56 \hphantom{1}[3.7]	\\
8	& 1192 [79.1]	& 8 \hphantom{1}[0.5]	& 22 \hphantom{1}[1.5]	\\
9	& 255 [16.9]	& 9 \hphantom{1}[0.6]	& 18 \hphantom{1}[1.2]	\\
10	& 131 \hphantom{1}[8.7]	& 0 \hphantom{1}[0.0]	& 13 \hphantom{1}[0.9]	\\
11	& 35 \hphantom{1}[2.3]	& 1 \hphantom{1}[0.1]	& 6 \hphantom{1}[0.4]	\\
12	& 11 \hphantom{1}[0.7]	& 0 \hphantom{1}[0.0]	& 11 \hphantom{1}[0.7]	\\
\end{tabular}
	\end{minipage}
	
		\vspace{3mm}
		
		\begin{center}
	(e) \textbf{Non-STEM}, $N_{\textrm{unique}} = 5099$
\end{center}

\begin{tabular}{c | r r r}
	\multirow{2}{*}{Term}	& \multicolumn{3}{c}{Number of Majors [\% of $N_{\textrm{unique}}$]} \\
	& Current	& Added	& Dropped	\\
	\hline
1	& 254 \hphantom{1}[5.0]	& 254 \hphantom{1}[5.0]	& 1 \hphantom{1}[0.0]	\\
2	& 1012 [19.8]	& 790 [15.5]	& 40 \hphantom{1}[0.8]	\\
3	& 2456 [48.2]	& 1487 [29.2]	& 51 \hphantom{1}[1.0]	\\
4	& 3527 [69.2]	& 1159 [22.7]	& 110 \hphantom{1}[2.2]	\\
5	& 4141 [81.2]	& 745 [14.6]	& 81 \hphantom{1}[1.6]	\\
6	& 4351 [85.3]	& 304 \hphantom{1}[6.0]	& 68 \hphantom{1}[1.3]	\\
7	& 4367 [85.6]	& 216 \hphantom{1}[4.2]	& 110 \hphantom{1}[2.2]	\\
8	& 4125 [80.9]	& 107 \hphantom{1}[2.1]	& 108 \hphantom{1}[2.1]	\\
9	& 796 [15.6]	& 45 \hphantom{1}[0.9]	& 67 \hphantom{1}[1.3]	\\
10	& 443 \hphantom{1}[8.7]	& 10 \hphantom{1}[0.2]	& 32 \hphantom{1}[0.6]	\\
11	& 114 \hphantom{1}[2.2]	& 4 \hphantom{1}[0.1]	& 14 \hphantom{1}[0.3]	\\
12	& 57 \hphantom{1}[1.1]	& 3 \hphantom{1}[0.1]	& 12 \hphantom{1}[0.2]	\\
\end{tabular}
	
	\caption{\label{table_adddrop_2}
		For each term from 1 to 12, the current number of declared majors (``Current'') is shown along with the number of majors who newly declared in that term (``Added'') and the number of former majors who dropped the major as of that term (``Dropped'').
		In square brackets next to each measure is the percentage of all unique students who declared that major (or cluster of majors as in (d) and (e)).
		The five sub-tables show this information for five different majors or clusters of majors: (a) biological sciences, (b) economics, (c) psychology, (d) other STEM disciplines not listed separately in Tables II and III, and (e) other non-STEM majors excluding psychology.
		}
\end{table*}

\clearpage

\section{Appendix B: Summary Tables of Declared Majors}
\label{appendix_B}

\begin{table*}[b]
    \centering

		\begin{tabular}{l | r r r r}
	\textbf{All Students}	& Unique	& Max Simultaneous	& Max Added		& Max Dropped	\\
	Major					& Majors 	& Majors [Term]		& Majors [Term]	& Majors [Term]	\\
	\hline
Biological Sciences	& 2211	& 1933 [5]	& 1141 [3]	& 87 [4]	\\
Chemistry	& 1063	& 919 [5]	& 706 [3]	& 59 [4]	\\
Computer Science	& 603	& 489 [7]	& 184 [5]	& 17 [7]	\\
Economics	& 934	& 742 [7]	& 267 [4]	& 51 [7]	\\
Engineering	& 3505	& 3164 [1]	& 3164 [1]	& 259 [2]	\\
Mathematics	& 367	& 280 [6]	& 102 [3]	& 26 [5]	\\
Physics and Astronomy	& 182	& 148 [4]	& 62 [3]	& 14 [3]	\\
Psychology	& 1767	& 1643 [7]	& 614 [4]	& 29 [6]	\\
Other STEM	& 1507	& 1315 [5]	& 631 [3]	& 56 [7]	\\
Non-STEM	& 5099	& 4367 [7]	& 1487 [3]	& 110 [7]	\\
\end{tabular}
		
	\caption{\label{table_appendix_all_main}
		Summary counts for all students.
		For each major, the total number of unique students is listed along with peak concurrent majors, added majors, and dropped majors, as well as the term in which the peak occurs in brackets.
		For example, in biological sciences, there were 2211 individual students in the sample who had ever declared the major.
		1933 of those students declared biological science majors in term 5 (peak term for concurrent majors), which is higher than the number of majors declared in any other term.
		Further, 1141 of those students added the major in term 3 (peak term for adding this major) and 87 of those students dropped the major in term 4 (peak term for dropping this major).
		}
\end{table*}

\begin{table*}[b]
    \centering

		\begin{tabular}{l | r r r r}
	\textbf{Asian Students}	& Unique	& Max Simultaneous	& Max Added		& Max Dropped	\\
	Major					& Majors 	& Majors [Term]		& Majors [Term]	& Majors [Term]	\\
	\hline
Biological Sciences	& 433	& 383 [5]	& 246 [3]	& 22 [6]	\\
Chemistry	& 115	& 95 [6]	& 64 [3]	& 10 [4]	\\
Computer Science	& 93	& 74 [8]	& 20 [5]	& 4 [7]	\\
Economics	& 185	& 153 [6]	& 53 [3]	& 14 [7]	\\
Engineering	& 295	& 255 [3]	& 255 [1]	& 16 [2]	\\
Mathematics	& 49	& 36 [6]	& 16 [3]	& 6 [4]	\\
Physics and Astronomy	& 11	& 9 [4]	& 5 [4]	& 2 [5]	\\
Psychology	& 209	& 191 [7]	& 70 [4]	& 6 [6]	\\
Other STEM	& 302	& 270 [5]	& 134 [3]	& 13 [6]	\\
Non-STEM	& 477	& 395 [6]	& 124 [3]	& 16 [7]	\\
\end{tabular}
				
	\caption{\label{table_appendix_asian_main}
		Summary counts for Asian students.
		For each major, the total number of unique Asian students is listed along with peak concurrent majors, added majors, and dropped majors, as well as the term in which the peak occurs in brackets.
		}
\end{table*}

\begin{table*}[b]
    \centering

		\begin{tabular}{l | r r r r}
	\textbf{URM Students}	& Unique	& Max Simultaneous	& Max Added		& Max Dropped	\\
	Major					& Majors 	& Majors [Term]		& Majors [Term]	& Majors [Term]	\\
	\hline
Biological Sciences	& 196	& 166 [6]	& 85 [3]	& 10 [4]	\\
Chemistry	& 86	& 70 [5]	& 55 [3]	& 5 [6]	\\
Computer Science	& 71	& 62 [8]	& 30 [5]	& 4 [11]	\\
Economics	& 96	& 72 [7]	& 31 [4]	& 7 [8]	\\
Engineering	& 344	& 310 [1]	& 310 [1]	& 28 [2]	\\
Mathematics	& 24	& 19 [8]	& 7 [3]	& 2 [6]	\\
Physics and Astronomy	& 17	& 15 [5]	& 11 [3]	& 3 [7]	\\
Psychology	& 267	& 243 [7]	& 85 [5]	& 8 [6]	\\
Other STEM	& 139	& 120 [5]	& 53 [3]	& 10 [5]	\\
Non-STEM	& 590	& 506 [7]	& 159 [3]	& 20 [8]	\\
\end{tabular}
				
	\caption{\label{table_appendix_urm_main}
		Summary counts for URM students.
		For each major, the total number of unique URM students is listed along with peak concurrent majors, added majors, and dropped majors, as well as the term in which the peak occurs in brackets.
		}
\end{table*}

\begin{table*}[b]
    \centering

		\begin{tabular}{l | r r r r}
	\textbf{White Students}	& Unique	& Max Simultaneous	& Max Added		& Max Dropped	\\
	Major					& Majors 	& Majors [Term]		& Majors [Term]	& Majors [Term]	\\
	\hline
Biological Sciences	& 1582	& 1384 [5]	& 810 [3]	& 61 [4]	\\
Chemistry	& 862	& 758 [5]	& 587 [3]	& 44 [4]	\\
Computer Science	& 439	& 363 [7]	& 134 [5]	& 12 [7]	\\
Economics	& 653	& 522 [7]	& 192 [4]	& 33 [7]	\\
Engineering	& 2866	& 2599 [1]	& 2599 [1]	& 215 [2]	\\
Mathematics	& 294	& 226 [6]	& 79 [4]	& 25 [5]	\\
Physics and Astronomy	& 154	& 126 [4]	& 50 [2]	& 12 [3]	\\
Psychology	& 1291	& 1209 [7]	& 463 [4]	& 22 [8]	\\
Other STEM	& 1066	& 925 [5]	& 444 [3]	& 42 [4]	\\
Non-STEM	& 4032	& 3470 [7]	& 1204 [3]	& 83 [4]	\\
\end{tabular}
				
	\caption{\label{table_appendix_white_main}
		Summary counts for White students.
		For each major, the total number of unique White students is listed along with peak concurrent majors, added majors, and dropped majors, as well as the term in which the peak occurs in brackets.
		}
\end{table*}

\clearpage

\section{Appendix C: Degrees Earned by Dropped Majors}
\label{appendix_C}

\begin{table*}[b]
    \centering
    
\begin{tabular}{l r | R{0.75cm} R{0.85cm} R{0.75cm} R{0.75cm} R{0.75cm} R{0.75cm} R{0.75cm} R{0.825cm} R{0.95cm} R{0.95cm} R{1cm}}
	\multicolumn{2}{c|}{\textbf{All Students}}		& \multicolumn{11}{c}{\% of $N_{\textrm{drop}}$ in a Given Major That Subsequently Earned Degree in Each Major} \\
								&					& & & & & & & & & Other & Non-& No \\
	Major						& $N_{\textrm{drop}}$	& Bio 
													& Chem 
													& CS
													& Econ
													& Engr
													& Math
													& Phys
													& Psych
													& STEM
													& STEM
													& Degree
													\\
	\hline
Bio			& 450	& 2.7	& 4.0	& 3.1	& 1.6	& 10.2	& 2.2	& 0.2	& 7.8	& 7.8	& 44.0	& 27.1	\\
Chem		& 218	& 10.6	& 1.4	& 4.6	& 1.8	& 11.9	& 1.8	& 0.9	& 5.0	& 7.8	& 38.1	& 26.1	\\
CS			& 80	& 5.0	& 0.0	& 1.2	& 2.5	& 2.5	& 3.8	& 1.2	& 2.5	& 2.5	& 26.2	& 55	\\
Econ		& 226	& 2.7	& 0.9	& 3.5	& 1.3	& 2.2	& 7.1	& 0.0	& 4.4	& 7.1	& 47.3	& 31.9	\\
Engr		& 790	& 4.2	& 2.7	& 8.2	& 5.3	& 0.9	& 2.4	& 0.8	& 2.4	& 4.9	& 29.9	& 46.1	\\
Math		& 122	& 10.7	& 5.7	& 4.9	& 15.6	& 5.7	& 4.1	& 1.6	& 7.4	& 19.7	& 20.5	& 25.4	\\
Phys		& 68	& 1.5	& 2.9	& 4.4	& 4.4	& 11.8	& 16.2	& 1.5	& 2.9	& 2.9	& 25.0	& 36.8	\\
Psych		& 154	& 1.9	& 0.0	& 3.2	& 1.3	& 0.6	& 0.0	& 0.6	& 9.7	& 2.6	& 35.7	& 51.3	\\
Other STEM	& 304	& 11.2	& 1.3	& 3.0	& 11.5	& 2.6	& 5.6	& 0.0	& 15.5	& 0.7	& 35.9	& 27.0	\\
Non-STEM	& 694	& 6.6	& 1.3	& 4.3	& 4.5	& 2.7	& 1.6	& 1.0	& 12.4	& 5.2	& 8.4	& 59.4	\\
\end{tabular}
	
	\caption{\label{table_appendix_all_supp}
		Trajectory of all students who dropped a major.
		For each major, the total number of students in the dataset who dropped that major ($N_{\textrm{drop}}$) is listed along with the percentage of $N_{\textrm{drop}}$ who ultimately earned a degree in each major or earned no degree.
		For example, there were 450 students who ever dropped their major in biological sciences.
		Of those 450 students, 2.7\% went on to earn a degree in biological sciences (i.e., they later declared that major again after dropping it at an earlier point).
		Similarly, 4.0\% of them earned a degree in chemistry, 3.1\% in computer science, 1.6\% in economics, 10.2\% in engineering, and so forth.
		Finally, 27.1\% of those 450 students that dropped a biological sciences major ultimately did not earn a degree from the university.
		}
\end{table*}

\begin{table*}[b]
    \centering
    
\begin{tabular}{l r | R{0.75cm} R{0.85cm} R{0.75cm} R{0.75cm} R{0.75cm} R{0.75cm} R{0.75cm} R{0.825cm} R{0.95cm} R{0.95cm} R{1cm}}
	\multicolumn{2}{c|}{\textbf{Asian Students}}	& \multicolumn{11}{c}{\% of $N_{\textrm{drop}}$ in a Given Major That Subsequently Earned Degree in Each Major} \\
								&					& & & & & & & & & Other & Non-& No \\
	Major						& $N_{\textrm{drop}}$	& Bio 
													& Chem 
													& CS
													& Econ
													& Engr
													& Math
													& Phys
													& Psych
													& STEM
													& STEM
													& Degree
													\\
	\hline
Bio			& 92	& 5.4	& 1.1	& 5.4	& 4.3	& 9.8	& 3.3	& 0.0	& 6.5	& 13.0	& 42.4	& 25.0	\\
Chem		& 26	& 11.5	& 3.8	& 7.7	& 0.0	& 11.5	& 0.0	& 0.0	& 7.7	& 11.5	& 30.8	& 26.9	\\
CS			& 13	& 7.7	& 0.0	& 0.0	& 7.7	& 0.0	& 0.0	& 0.0	& 7.7	& 0.0	& 23.1	& 61.5	\\
Econ		& 49	& 2.0	& 4.1	& 4.1	& 0.0	& 2.0	& 6.1	& 0.0	& 10.2	& 14.3	& 28.6	& 36.7	\\
Engr		& 66	& 3.0	& 3.0	& 13.6	& 9.1	& 0.0	& 1.5	& 1.5	& 3.0	& 7.6	& 24.2	& 47.0	\\
Math		& 16	& 6.2	& 0.0	& 6.2	& 18.8	& 6.2	& 0.0	& 0.0	& 12.5	& 31.2	& 18.8	& 31.2	\\
Phys		& 5	& 0.0	& 0.0	& 0.0	& 0.0	& 20.0	& 20.0	& 0.0	& 0.0	& 0.0	& 40.0	& 20.0	\\
Psych		& 14	& 7.1	& 0.0	& 0.0	& 0.0	& 0.0	& 0.0	& 0.0	& 7.1	& 0.0	& 50.0	& 35.7	\\
Other STEM	& 55	& 10.9	& 1.8	& 5.5	& 9.1	& 1.8	& 3.6	& 0.0	& 18.2	& 1.8	& 30.9	& 29.1	\\
Non-STEM	& 62	& 12.9	& 1.6	& 3.2	& 1.6	& 3.2	& 0.0	& 0.0	& 16.1	& 8.1	& 4.8	& 53.2	\\
\end{tabular}
	
	\caption{\label{table_appendix_asian_supp}
		Trajectory of all Asian students who dropped a major.
		For each major, the total number of Asian students in the dataset who dropped that major ($N_{\textrm{drop}}$) is listed along with the percentage of $N_{\textrm{drop}}$ who ultimately earned a degree in each major or earned no degree.
		}
\end{table*}

\begin{table*}[b]
    \centering
    
\begin{tabular}{l r | R{0.75cm} R{0.85cm} R{0.75cm} R{0.75cm} R{0.75cm} R{0.75cm} R{0.75cm} R{0.825cm} R{0.95cm} R{0.95cm} R{1cm}}
	\multicolumn{2}{c|}{\textbf{URM Students}}		& \multicolumn{11}{c}{\% of $N_{\textrm{drop}}$ in a Given Major That Subsequently Earned Degree in Each Major} \\
								&					& & & & & & & & & Other & Non-& No \\
	Major						& $N_{\textrm{drop}}$	& Bio 
													& Chem 
													& CS
													& Econ
													& Engr
													& Math
													& Phys
													& Psych
													& STEM
													& STEM
													& Degree
													\\
	\hline
Bio			& 52	& 1.9	& 1.9	& 0.0	& 0.0	& 1.9	& 0.0	& 0.0	& 9.6	& 0.0	& 53.8	& 30.8	\\
Chem		& 20	& 5.0	& 0.0	& 10.0	& 5.0	& 20.0	& 10.0	& 0.0	& 5.0	& 5.0	& 45.0	& 15.0	\\
CS			& 7	& 14.3	& 0.0	& 0.0	& 0.0	& 0.0	& 0.0	& 0.0	& 0.0	& 0.0	& 28.6	& 57.1	\\
Econ		& 33	& 0.0	& 0.0	& 6.1	& 0.0	& 6.1	& 6.1	& 0.0	& 6.1	& 6.1	& 45.5	& 36.4	\\
Engr		& 98	& 3.1	& 1.0	& 7.1	& 1.0	& 0.0	& 3.1	& 0.0	& 2.0	& 3.1	& 28.6	& 54.1	\\
Math		& 6	& 0.0	& 0.0	& 0.0	& 33.3	& 16.7	& 16.7	& 0.0	& 0.0	& 16.7	& 16.7	& 33.3	\\
Phys		& 11	& 9.1	& 0.0	& 0.0	& 9.1	& 9.1	& 18.2	& 9.1	& 9.1	& 0.0	& 27.3	& 18.2	\\
Psych		& 28	& 3.6	& 0.0	& 7.1	& 0.0	& 0.0	& 0.0	& 0.0	& 7.1	& 0.0	& 28.6	& 64.3	\\
Other STEM	& 36	& 11.1	& 2.8	& 0.0	& 5.6	& 0.0	& 5.6	& 0.0	& 8.3	& 0.0	& 47.2	& 27.8	\\
Non-STEM	& 107	& 8.4	& 0.0	& 1.9	& 2.8	& 0.9	& 0.0	& 0.9	& 8.4	& 2.8	& 8.4	& 71.0	\\
\end{tabular}
	
	\caption{\label{table_appendix_urm_supp}
		Trajectory of all URM students who dropped a major.
		For each major, the total number of URM students in the dataset who dropped that major ($N_{\textrm{drop}}$) is listed along with the percentage of $N_{\textrm{drop}}$ who ultimately earned a degree in each major or earned no degree.
		}
\end{table*}

\begin{table*}[b]
    \centering
    
\begin{tabular}{l r | R{0.75cm} R{0.85cm} R{0.75cm} R{0.75cm} R{0.75cm} R{0.75cm} R{0.75cm} R{0.825cm} R{0.95cm} R{0.95cm} R{1cm}}
	\multicolumn{2}{c|}{\textbf{White Students}} 	& \multicolumn{11}{c}{\% of $N_{\textrm{drop}}$ in a Given Major That Subsequently Earned Degree in Each Major} \\
								&					& & & & & & & & & Other & Non-& No \\
	Major						& $N_{\textrm{drop}}$	& Bio 
													& Chem 
													& CS
													& Econ
													& Engr
													& Math
													& Phys
													& Psych
													& STEM
													& STEM
													& Degree
													\\
	\hline
Bio			& 306	& 2.0	& 5.2	& 2.9	& 1.0	& 11.8	& 2.3	& 0.3	& 7.8	& 7.5	& 42.8	& 27.1	\\
Chem		& 172	& 11.0	& 1.2	& 3.5	& 1.7	& 11.0	& 1.2	& 1.2	& 4.7	& 7.6	& 38.4	& 27.3	\\
CS			& 60	& 3.3	& 0.0	& 1.7	& 1.7	& 3.3	& 5.0	& 1.7	& 1.7	& 3.3	& 26.7	& 53.3	\\
Econ		& 144	& 3.5	& 0.0	& 2.8	& 2.1	& 1.4	& 7.6	& 0.0	& 2.1	& 4.9	& 54.2	& 29.2	\\
Engr		& 626	& 4.5	& 2.9	& 7.8	& 5.6	& 1.1	& 2.4	& 0.8	& 2.4	& 5.0	& 30.7	& 44.7	\\
Math		& 100	& 12.0	& 7.0	& 5.0	& 14	& 5.0	& 4.0	& 2.0	& 7.0	& 18.0	& 21.0	& 24.0	\\
Phys		& 52	& 0.0	& 3.8	& 5.8	& 3.8	& 11.5	& 15.4	& 0.0	& 1.9	& 3.8	& 23.1	& 42.3	\\
Psych		& 112	& 0.9	& 0.0	& 2.7	& 1.8	& 0.9	& 0.0	& 0.9	& 10.7	& 3.6	& 35.7	& 50.0	\\
Other STEM	& 213	& 11.3	& 0.9	& 2.8	& 13.1	& 3.3	& 6.1	& 0.0	& 16.0	& 0.5	& 35.2	& 26.3	\\
Non-STEM	& 525	& 5.5	& 1.5	& 5.0	& 5.1	& 3.0	& 2.1	& 1.1	& 12.8	& 5.3	& 8.8	& 57.7	\\
\end{tabular}
	
	\caption{\label{table_appendix_white_supp}
		Trajectory of all White students who dropped a major.
		For each major, the total number of White students in the dataset who dropped that major ($N_{\textrm{drop}}$) is listed along with the percentage of $N_{\textrm{drop}}$ who ultimately earned a degree in each major or earned no degree.
		}

\end{table*}

\clearpage

\section{Appendix D: GPA of Those That Drop vs. Those That Earn a Degree in a Major by Racial/Ethnic Group}
\label{appendix_D}

\begin{figure*}[b]
    \centering
    
	\includegraphics[width=0.95\textwidth]{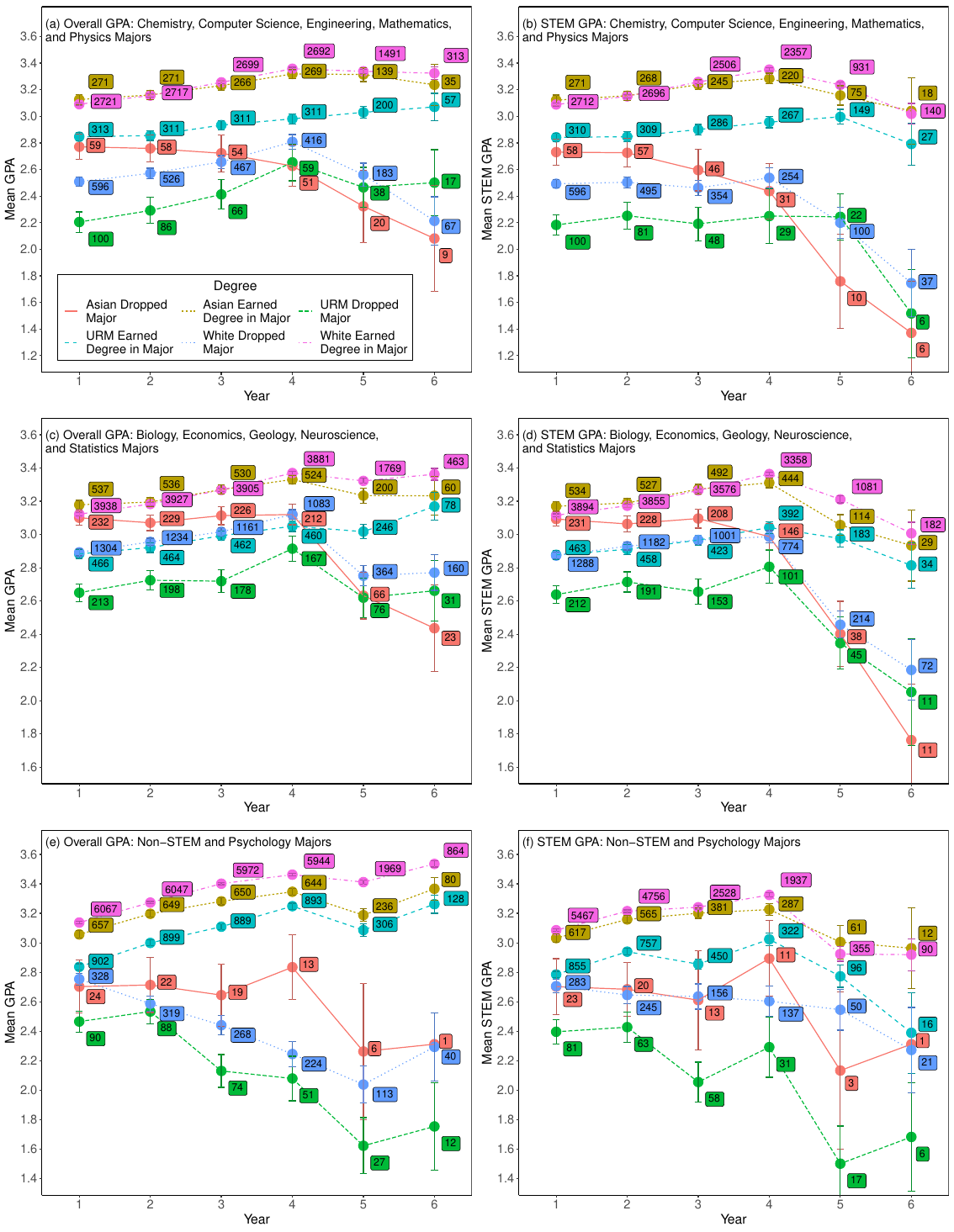}
	
	\caption{\label{figure_race_gpa_major_all}
	GPA and STEM GPA over time by racial/ethnic group.
	Majors are divided into three groupings: (a) and (b) chemistry, computer science, engineering, mathematics, and physics; (c) and (d) biology, economics, geology, neuroscience, and statistics; and (e) and (f) non-STEM including psychology.
	GPA in all courses -- (a), (c), and (e) -- and in only STEM courses -- (b), (d), and (f) -- are calculated separately for four categories of students that declared at least one of the majors in each group: Asian, URM, and White students that ultimately earned a degree in that group of majors and those that dropped that group of majors, as indicated by the legend within (a).
	For each group, the mean GPA is plotted along with its standard error, with the sample size listed above each point and guides to the eye connecting the points.
	Note that the vertical axis scale differs among the different plots.
	}
\end{figure*}

\end{document}